\documentclass[11pt]{article}
\usepackage{amsmath, amssymb, graphics}

\usepackage{graphics, psfrag}
\usepackage{graphicx}

\def\id{\protect{{1 \kern-.28em {\rm l}}}}

\def\be{\begin{eqnarray}}
\def\ee{\end{eqnarray}}
\def\p{{\partial}}
\def\nn{\nonumber}
\makeatletter
\renewcommand\section{\@startsection {section}{1}{\z@}%
								   {-3.5ex \@plus -1ex \@minus -.2ex}%
								   {2.3ex \@plus.2ex}%
								   {\normalfont\large\bfseries}}
\renewcommand\subsection{\@startsection{subsection}{2}{\z@}%
								   {-3.25ex\@plus -1ex \@minus -.2ex}%
								   {1.5ex \@plus .2ex}%
								   {\normalfont\normalsize\bfseries}}

\makeatother

\newcommand{\mathsym}[1]{{}}
\def\k{\kappa}

\def\p{{\partial}}
\def\nn{\nonumber}

\def\dalemb#1#2{{\vbox{\hrule height .#2pt
		\hbox{\vrule width.#2pt height#1pt \kern#1pt
				\vrule width.#2pt}
		\hrule height.#2pt}}}

\let\a=\alpha \let\b=\beta \let\g=\gamma \let\d=\delta \let\e=\epsilon
\let\z=\zeta  \let\th=\theta  \let\k=\kappa
\let\l=\lambda \let\m=\mu  \let\x=\xi \let\p=\pi %\let\r=\rho
\let\s=\sigma \let\t=\tau   \let\c=\chi 
\let\vp=\varphi \let\vep=\varepsilon
\let\w=\omega	   \let\G=\Gamma \let\D=\Delta \let\Th=\Theta \let\L=\Lambda
 \let\P=\Pi \let\S=\Sigma  
\let\C=\Chi \let\W=\Omega
\let\la=\label \let\ci=\cite 
  
\def\nn{\nonumber} \def\bd{\begin{document}} \def\ed{\end{document}}
\def\ds{\documentstyle} \let\fr=\frac \let\bl=\bigl \let\br=\bigr
\let\Br=\Bigr \let\Bl=\Bigl
\let\bm=\bibitem
\let\na=\nabla
\def\tU{{\widetilde U}}
\let\pa=\partial \let\ov=\overline
\def\ie{{\it i.e.\ }}
\def\ba{\begin{array}}
\def\ea{\end{array}}
\def\ft#1#2{{\textstyle{{\scriptstyle #1}\over {\scriptstyle #2}}}}
\def\fft#1#2{{#1 \over #2}}
\def\F#1#2{{ F_{#1}^{(#2)} }}
\def\cF#1#2{{ {\cal F}_{#1}^{(#2)} }}

\def\R{{\bf R}}
\def\sst#1{{\scriptscriptstyle #1}}
\def\oneone{\rlap 1\mkern4mu{\rm l}}
\def\e7{E_{7(+7)}}
\def\td{\tilde}
\def\wtd{\widetilde}
\def\im{{\rm i}}
\newcommand{\ho}[1]{$\, ^{#1}$}
\newcommand{\hoch}[1]{$\, ^{#1}$}
\newcommand{\bea}{\begin{eqnarray}}
\newcommand{\eea}{\end{eqnarray}}
\newcommand{\ra}{\rightarrow}
\newcommand{\lra}{\longrightarrow}
\newcommand{\Lra}{\Leftrightarrow}
\newcommand{\ap}{\alpha^\prime}
\newcommand{\bp}{\tilde \beta^\prime}
\newcommand{\cB}{{\cal B}}
\newcommand{\cO}{{\cal O}}
\newcommand{\vecx}{\vec{x}}
\newcommand{\vecy}{\vec{y}}
\newcommand{\vecp}{\vec{p}}
\newcommand{\vecq}{\vec{q}}
\newcommand{\tr}{{\rm tr} }
\newcommand{\Tr}{{\rm Tr} }

\newcommand{\cL}{{\cal L}}
\newcommand{\cA}{{\cal A}}
\newcommand{\cD}{{\cal D}}
\def\sst#1{{\scriptscriptstyle #1}}

\def\ve{\varepsilon}
\def\vf{\varphi}
\def\F{\Phi}
\def\wg{\wedge}

\newcommand{\wt}{\widetilde}
\newcommand{\oh}[1]{{\cal O}( #1 )}
\newcommand{\largeoh}[1]{{\cal O}\left( #1 \right)}
\def \foot {\footnote}
\def \bi{\bibitem}

\def \tr {{\rm tr}}
\def \ha {{1 \over 2}}
\def \td {\tilde}
\def \ci{\cite}
\def \N {{\mathcal N}}
\def \const {{\rm const}}
\def \ss {\sum\limits_{i=1}^3 }
\def \t {\tau}
\def\S{{\mathcal S} }
\def \nn {\nu}
\def \XX {{\rm X}}

\def \lra {\leftrightarrow}
\def \vom {{\bar \omega}}
\def \E {{\mathcal	E}} \def \J {{\mathcal	J}}
\def \YY {{\rm Y}}

\def \d {\del}
\def \rJ {{J}}
\def \sms {sigma models\ }
\def \sm {sigma model\ }
\def \L {\Lambda}
\def \gl {\ell}
\def \tr {{\rm tr\ }}
\def\z{\zeta}
\def\zi{\zeta_1}
\def\zii{\zeta_2}
\def\K{\mbox{K}}
\def\eE{\mbox{E}}	\def \vt {\vartheta}
\def \vr {\varrho}
\def \wup {w}

\def\dg{\dagger}
\def\a{\alpha}
\def\b{\beta}
\def\e{\varepsilon}
\def\p{\phi}
\def\ap{\alpha^\prime}
\def\I{{\cal I}}

\def\R{{\bf R}}
\def\Z{{\bf Z}}
\def\C{{\bf C}}
\def\P{{\bf P}}
\def\xb{{\bar X}}
\def\Tr{{\rm  Tr}}
\def\tr{{\rm  tr}}

\def \bs {\bar \s}
\def \btau {\bar \tau}

\def \del{\partial}
\def \a {\alpha}
\def \aa {{\a'}}
\def\g{\gamma}
\def\s{\sigma}
\def\z{\zeta}
\def\zi{\zeta_1}
\def\zii{\zeta_2}
\def\ov{\over}

\def\I{{\cal I}}
\def\J{{\mathcal J}}
\def \ok {{1\ov \k}}
\def\LL{{\mathcal L }}
\def \jL {{J}}
\def \om {\omega}
\def \cL {{\mathcal L}} \def \cH {{\mathcal H}}
\def\E{{\mathcal E}}
\def\b{\beta}
\def\l{\lambda}
\def\eps{\epsilon}
\def\vep{\varepsilon}
\def \De {{\mathcal D}}

 \def \cV {{\cal V}}
\def  \Jt {	 {J}_{\rm tot}	  }

\def \k {\kappa}
\def\foot{\footnote}
\def \four{{\textstyle {1\ov 4}}}
 \def \third { \textstyle {1\ov 3
}}
\def\det{\hbox{det}}
\def \ci {\cite}

\def \foot {\footnote}
\def \bi{\bibitem}

\def \tr {{\rm tr}}
\def \ha {{1 \over 2}}
\def \tid {\tilde}
\def \vv {{\rm v}}
\def \tl {{\tilde \l}}
\def \XX {{\rm X}}
\def \ta {{\tilde \a}}
\def \fo { {1\ov 4}}
\def \ep {\epsilon}
\def \inti {{\int^{2\pi}_0 {d \sigma \ov 2 \pi}}}

\def \d {\partial}
\def \K {{\rm S}}
\def \el {\ell}
\def \Tr {{\rm Tr}}
\def \P {\Phi}
\def \l	 {\lambda}
\def \tl {{\tilde \l}}
\def \bl {{\tilde \l}}
\def \const {{\rm const}}
\def \V {v}

\def \bv {v^*}
\def \vv {{\rm v}}
\def \LL {{\mathcal L}}
\newcommand{\PV}[1]{P_{\!\!_{V_{#1}}}}

\def \bL {\ell}
\def \M {{\mathcal M}}
\def \N {{\mathcal N}}
\def \S {{\rm S}}
\def \vn {\vec n}
\def \tl {\td \l}
\def \td {\tilde}
\def \Prod {\Pi}
\def \O {{\mathcal O}}
\def \Q {{\rm  Q}}
\def \D {\Delta}
\def \N {{\mathcal N}}
\def\tN{{\tilde N}}

\def \m {\mu}
\def \vs {\vec \s}
\def \ie {i.e.}

\def \cD {{\cal D}}

\def  \le  {\l_{\rm eff}}

\def\as{{\a}}
\newcommand{\bra}[1]{\mbox{$\langle #1 |$}}
\newcommand{\ket}[1]{\mbox{$| #1 \rangle$}}

\def\thb{\bar{\theta}}
\def\Thb{\bar{\Theta}}
\def\barp{\bar{p}}
\def\barq{\bar{q}}
\def\barc{\bar{c}}
\def\bard{\bar{d}}
\def\e{\epsilon}

\def \bi{\bibitem}
\def \la {\label}

\def \l {\lambda}
\def\foot{\footnote}
\def \tl  {{\tilde \l}}
\def \sql {{\sqrt \l}}
\def \adss {$AdS_5 \times S^5$\ }
\newcommand{\rf}[1]{(\ref{#1})}
\def \ov {\over}

\def\th{\theta}
\def\Th{\Theta}
\def\vth{\vartheta}

\def\vth{\vartheta}
\def\ra{\rightarrow}
\def\N{{\cal N}}
\def\F{{\cal F}}
\def\cc{\circ}
\def\eqv{\equiv}

\def\ni{\noindent}
\def \ha{{1\ov 2}}
\def \bw {{\rm w}}

\def\r{{\rm r}}

\def \cT {{\cal T}}
\def \no {\nonumber}
\def \J {\mathcal{J}}
\def \del {\partial}

\def \bps {{\bar \psi}}
\def \sqbl {\sqrt{\bar \lambda}}

\def\dF{\dot{F}}
\def\dG{\dot{G}}
\def\df{\dot{f}}
\def \E {{\cal E}}

\def \S {{\cal S}}
\def \J {{\cal J}}

\def\ms{\mathcal{S}}
\def\mj{\mathcal{J}}
\def\soj{\fr{\ms}{\mj}}
\def \R {{\bf R}}
\def \tH {\widetilde H}
\def \bE {\bar E}
\def \x {{\cal X}}

\def \hV {{\hat V}}

 \def \bb {\bar \beta}

\def \bi{\bibitem}
\def \la {\label}

\def \l {\lambda}
\def\foot{\footnote}
\def \tl  {{\tilde \l}}
\def \sql {{\sqrt \l}}
\def \sqtl {{\sqrt {\tilde \l}}}

\def \HH {{\rm E}}
\def \cS {{\cal S}}
\def \cL {{\cal L}}
\def \ka {\kappa}

\def \adss {$AdS_5 \times S^5$\ }
\def \arccot {{\rm arccot}}

\def \D {\Delta}
 \def \t {\tau}
 \def \p {\phi}
 \def \r {\rho}
 \def \rN {{\rm N}}
 \def\tw{{\tilde w}}
 \def\hJ{{J}}
 \def\hw {{w}}
 \def\hl{{\lambda}}
 \def\hth{{\theta}}
 \def\NN{{\cal N}}
 \def \bv {{ \bar w}}
\def \vn {{\vec n}}
\newcommand{\sfrac}[2]{{\textstyle\frac{#1}{#2}}}
\def \bl {{ \bar \lambda}}
\def \bp {{\bar p}}
\def \bu {{\bar u}}
\def \sha {\sfrac{1}{2}}
\def \ov {\over}
\def \vl { \vec \ell}
\def \varpi {{\rm w}}
\def \OO {{\cal O}}
\def \bG {\bar \G}

\def \c {\gamma}

\def \rE {{\rm E}}

\def \rK {{\rm K}}
\def \ve {\varepsilon}
\def \pa{\partial}
\def \I {{\cal I}}
\def \LL {{\cal L}}
\def \ep {\epsilon}
\def \R {{\rm R}}
\def \tilt {{\tilde t}}

\def\pic #1#2{\hbox{\lower#1pt\hbox{~\mbox{\epsfxsize=20truemm \epsffile{#2}}}}}

\def\pic #1#2#3{\hbox{\lower#1pt\hbox{~\mbox{\includegraphics[scale=#3]{#2}}}}}

\def \bt {\bar\theta}
\def \te {\theta}

\def \cc {{\rm f}}
\def \d {\delta}

\def \cL {{\cal L}}
\def \S	 {{\cal S}}
\def \pp {{q}}
\def \vt {\vartheta}
\def \mm {{\cal	 \ell}}
\def \Z {{\cal Z}}
\def \pa {\partial}
\def \C {{\cal C}}
\def \be {\bea}
\def \ee {\eea}
\def \c {\gamma}  \def \d {\delta}
\def \eps {\epsilon}

\def \bp {\begin{pmatrix}}	\def \ep {\end{pmatrix}}

 \def \T {{\cal T}}

\def \bp {\begin{pmatrix}}	\def \epm {\end{pmatrix}}
\def \ha {{\textstyle{1 \ov 2}}}
\def \r {\rho}
\def \S {{\cal S}}
\def \z {\chi}

\def \sn {{\rm sn}}
\def \dn {{\rm dn}}

\def \rr {\r_-}

\def\ess{{\rm s}}
\def \nut {\tilde \nu}
\def\nuhatS{{{\hat\nu}_{S}}}

\def \const { {\rm const} }
\def \k {\kappa}
\def \t {\tau}
\def \acosh {{\rm arccosh \,}}
\def \AdSS {${\rm AdS}_5 \times {\rm S}^5 \ $}
\def \adss {${\rm AdS}_5 \times {\rm S}^5 $}
\def \AdS {${\rm AdS}_5 \ $}
\def \ads {${\rm AdS}_5 $}
\def \size {\displaystyle}
\def \bei {\begin{itemize}}
\def \eei {\end{itemize}}
\def \RS {$R_t \times S^5$}
\def \fix {{\rm fixed}}

\begin{document}

%%%%%%%%%%%%%%%%%%%%%%%%%%%%%%%%%%%%%%

\overfullrule=0pt
\parskip=2pt
\parindent=12pt
\headheight=0in \headsep=0in \topmargin=0in \oddsidemargin=0in

\vspace{ -3cm}
\thispagestyle{empty}
\vspace{-1cm}

\rightline{Imperial-TP-EM-2010-2}

\begin{center}
\vspace{1cm}
{\Large\bf

On classical string solutions in \AdSS

%\vspace{1.2cm}
\vspace{0.4cm}

	}

\vspace{.2cm}

E. M. Murchikova\footnote{{\bf e-mail}: e.murchikova@imperial.ac.uk}

\vskip 0.6cm

{\em
Blackett Laboratory, Imperial College,
London SW7 2AZ, U.K.
\\
\vskip 0.08cm
\vskip 0.08cm SINP, Moscow State University, Moscow, 119991, Russia
 }

\vspace{.2cm}

\end{center}

\begin{abstract}
 %%%%%%%%%%%%%%%%%%%%%%%%%%%%%%%%%

We discuss some new simple closed bosonic string  solutions in \adss \
that may be of interest in the context of AdS/CFT duality.
In the first part of  this work	 we consider solutions
with two spins	 ($S_1,S_2$) in \ads.
Starting from the flat-space solutions and using
 perturbation theory in the	 curvature of \AdS
space, we construct	 leading terms in the small two-spin   solution.
We find	 corrections to the leading Regge term in
the classical string energy	 and uncover a	discontinuity
in the spectrum for	 certain type of a solution.
We then	  analyze the connection between
small-spin and large-spin limits of string solutions in \ads.
We show that the $S_1=S_2$ solution in  \AdS found in earlier papers
admits both limits
only in the simplest  cases of the folded and rigid circular strings.
In the	second part of the paper
 we construct a new class of
 chiral solutions in \RS \ for which embedding coordinates of $S^5$ satisfy
the linear Laplace  equations. They
 generalize the previously studied rigid string solutions.
 We study in detail a simple nontrivial  example.

%%%%%%%%%%%%%%%%%%%%%%%%%%%%%%%%%
\end{abstract}

\newpage
\setcounter{equation}{0}
\setcounter{footnote}{0}
\setcounter{section}{0}

\section{Introduction}

\renewcommand{\theequation}{1.\arabic{equation}}
 \setcounter{equation}{0}

Semiclassical string solutions is a	 useful tool
for probing AdS/CFT correspondence \cite{bmn,gkp,ft1,ft2}.
In the closed-string sector, AdS energy of a closed
string expressed in terms of spins\foot{
The generic states of bosonic string in \AdSS may be labeled by the values of
three $SO(2,4)$ Cartan generators $(E,S_1,S_2)$ and three $SO(6)$ Cartan generators
$(J_1,J_2,J_3).$ We will be interested in ``spinning'' string solutions
that have nonzero value of these charges.} and string tension
$T=\frac{\sqrt{\l}}{2\pi}$, i.e.
%(or t'Hooft coupling $\l$)
$E(S_i, J_m; \l)$,	gives the strong-coupling limit of
the scaling dimension of the corresponding gauge-theory
operator  (see, e.g., \cite{tov, ple}).
 Also, in the open string sector, solutions
ending at the boundary of \AdS describe the strong
coupling limit of the associated Wilson loops
 \cite{19,ar}.

In this paper we present  some	new	 classical solutions
for a closed bosonic string in the \adss.

We shall first consider	 strings with two spins in \AdS part of \adss.
A natural ansatz for describing	  a rigid ``rotating'' string solution is
\cite{ft2} ($0 \leq \s < 2\pi$):
\begin{equation}
\label{InanC}
\ba{c}
\size
Y_0 + i Y_5 = y_0(\s) \ e^{i \kappa \tau} \ , \qquad %\\
Y_1 + i Y_2 = y_1(\s) \ e^{i \om_1 \tau} \ , \qquad % \\
Y_3 + i Y_4 = y_2(\s) \ e^{i \om_2 \tau} \ , \\
\size
\k , \ \w_1, \ \w_2 = \const \ .
\ea
\end{equation}
Here $Y_P$ are embedding coordinates
of $R^{2,4}$ with the metric
$\eta_{PQ}=(-1,+1,+1,+1,+1,-1);$ $Y_P Y^P = -1.$
 A general approach to finding such rigid string solutions was
 developed in \ci{afrt}. Using the reduction of the conformal-gauge string	sigma model
 to the 1D	 Neumann  integrable  model, one finds that the equations for $y_0, \ y_1, \ y_2$
  are those of a harmonic oscillator
 constrained to move on a 2d hyperboloid --- an integrable system with two integrals of motion
 $b_1,b_2$	with
  $b_1+b_2 = \k^2 + \om^2_1 + \om_2^2$.
In general, the solutions are expressed in terms of hyperelliptic functions and thus
are not easy  to analyze.
There are few special cases when they simplify --- when
the hyperelliptic surface degenerates into an elliptic one	 and $y_a(\s)$
can be	expressed  in terms of
the standard elliptic  functions. Two such  cases,  $\w_1 = \w_2,$
corresponding to $S_1 = S_2$ solution, and its boosted analog  with $\k = \w_2 \neq \w_1$
were studied in \cite{TirTs}.
 The existence of simple but  more general solution with
two	 unequal  spins is	an open question.
Recent
study of $\N = 4$ SYM states dual  to  minimal energy spinning string configuration
with two spins ($S_1, S_2$) with $\frac{S_1}{S_2}$ fixed  using
  the asymptotic Bethe ansatz (ABA)
\cite{rej} suggests that such simple solution might indeed exist.
In the	large-spin	limit  the energy  of the $S_1=S_2$ solution \cite{TirTs}
matched the	 strong-coupling ABA   result of \cite{rej}.

Aiming at a better understanding of two-spin solutions in \ads,
here we first study the case of small strings or small-spin limit. Starting from
the flat-space case, in which the general two-spin solutions
are	\cite{afrt}
\begin{equation}\la{Inflat}
\ba{c}
\size %t = \k \t, \qquad
\k^2 =	 n^2_1 a_1^2 + n^2_2 a_2^2	\\
\size {y}_{1}	 ^{\rm flat} = a_1	   \sin (n_1 \s) \qquad	 \size {y}_{2}^{\rm flat}	  = \ a_2 \sin [n_2 (\s + \s_0) ]	  \\
\size \w_1=n_1,
\qquad \w_2=n_2, \qquad n_i = {\rm integer} \ea
\end{equation}
and performing
perturbation with respect to the curvature of \ads, we
find the corrections to
the flat-space	expression for the	classical energy $E(S_1,S_2;\l).$
We uncover a discontinuity
in the spectrum of classical strings with equal and unequal winding numbers
in the $Y_1Y_2$ and $Y_3Y_4$ planes ($n_1$ and $n_2$).
It	may indicate that there are deep differences between solutions
 with $n_1 \neq n_2$ and more symmetrical ones with $n_1 = n_2.$
 We then investigate the connection between
small- (flat-space) and large-spin limits of
two-spin string solutions in \ads.
In the particular cases	 of $\w_1 = \w_2$ and $\k=\w_2 \neq \w_1$
the general solutions in \AdS were found in \ci{TirTs}.
It was discussed there,
for the $\k=\w_2 \neq \w_1$ case, that strings which admit a large-spin limit do
not have the small-spin one and vice-versa.
For $\w_1 = \w_2,$ we find that apart from the
trivial cases of folded and circular strings, the general
rigid solution with $S_1=S_2$ in \AdS admitting the large-spin limit does not have
a  small-spin limit. For more general two-spin
solutions, having both limits might	 still be possible.

In the second part of the paper we consider another
simple class of
string solutions --- chiral solutions in \RS. Such solutions
obey an additional constraint
\begin{equation}
   \partial_+ X_M \partial_- X_M=0,
\end{equation}
where $X_M$ are embedding coordinates
of $R^6$ with the Euclidean metric $\delta_{MN};$ $X_M X_M = 1$
 and
 $\partial_{\pm}=\frac{\partial}{\partial \sigma_{\pm}}= \frac{1}{2}\left(
\frac{\partial}{\partial \tau} \pm \frac{\partial}{\partial
\sigma} \right),$\ \ \	$\s_\pm = \t \pm \s.$
Then the classical string equations in the conformal gauge become
\be
\del_+ \del_- X_M =0.
\ee
The simplest solution of this kind	is	\cite{art},
\be\la{Inc}
 Y_0 + i Y_5 = e^{i \k \t} \ , \quad
 X_1 + i X_2 = a_1 e^{ i m_1 \s_\pm } \ , \quad
 X_3 + i X_4 = a_2 e^{ i m_2 \s_\pm } \ , \quad
 X_5 + i X_6 = a_3 e^{ i m_3 \s_\mp } \ ,
\ee
where $\ss a_i = 1$ and $m_i$ are integers. It was recently used
in	\cite{ko}  as a	  model of a quantum string state
with ``small'' quantum numbers.
We expect that more general	 chiral solutions
may also find  useful applications.

Here we	 consider the following	 ansatz
\begin{equation}\label{Inchan}
%	\displaystyle y_0=Y_5+iY_0=e^{i \kappa \tau} \\
   \displaystyle X_1 + i X_2 = a_1 e^{i F_1(\sigma_+) }, \quad
	  \displaystyle X_3 + i X_4 = a_2 e^{i F_2(\sigma_+ )}, \quad
   \displaystyle X_5 + i X_6 = a_3 e^{i F_3(\sigma_-) } ,
\end{equation}
and obtain the general solution for the functions
 $F_i(\s_\pm)$.
%\begin{equation}\label{InFi}
%F_{i}(\s_\pm)=m_{i}\s_{\pm} + \sum\limits_n {f_n^{(i)} \cos(n\s_{\pm}) + g_n^{(i)} \sin(n\s_{\pm}) } \ \qquad m_i = {\rm integer} ;
%\end{equation}
A particular simple nontrivial case
\be\la{InF}
F_1(\s_+) = \a \cos n\sigma_+ \ , \qquad
F_2(\s_+) = \a \sin n\sigma_+ \ , \qquad
F_3(\s_-) = m \s_-
\ee
we analyze in detail. It reduces to \rf{Inchan} in the limit $n \to 0.$
Note that  chiral solutions treat
$\t$ and $\s$  on an equal footing,
i.e.  nontrivial dependence on $\tau$ implies that the
shape of the string is not rigid, in general, so such solutions are similar to ``pulsating'' ones.

\vspace{20pt}

The rest of the paper is organized as follows.
In section 2 we discuss	 basics of
bosonic string solutions in \adss: action in the conformal
gauge, equations of motion, etc.
Section 3 is dedicated to small-string solutions in \ads.
In section 4 we consider the relation between string solutions admitting small- and
large-spin limits in \ads. In particular, we  discuss the
small-spin limit of exact solutions with two equal	spins.
Section 5 is devoted to the chiral solutions in \RS.
In	Appendix  A	 we give an overview of circular and
folded string solutions in \ads. In Appendix B curvature corrections to the
folded string solution displaced from the center of \ads \ are discussed.
In	Appendixes C and D	we present technical details of the calculation of
spins for chiral solutions corresponding to \rf{InF}.
%(Appendixes C and D).

\section{Closed bosonic string in \AdSS}

\renewcommand{\theequation}{2.\arabic{equation}}
 \setcounter{equation}{0}

We will be interested in the classical bosonic solutions for a closed string
in \AdSS 
\begin{equation}\label{1}
	I_B=\frac{1}{2}
		T \int d\tau \int \limits^{2\pi}_0 d\sigma (L_{AdS}+L_S),
	\qquad T=\frac{R^2}{2\pi \alpha'}=\frac{\sqrt{\lambda}}{2\pi},
\end{equation}
where
\begin{equation}\label{1.1}
	L_{AdS}=-\partial_a Y_P \partial^a Y^P-\tilde{\Lambda} (Y_P Y^P+1) ,
	\qquad
	L_{S}=-\partial_a X_M \partial^a X_M+\Lambda (X_M X_M-1).
\end{equation}
Here $X_M, \, M=1,...,6$ and $Y_P, \, P=0,...,5$ are embedding coordinates
of $R^6$ with the Euclidean metric $\delta_{MN}$ in $L_S$ and of $R^{2,4}$ with
$\eta_{PQ}=(-1,+1,+1,+1,+1,-1)$ in $L_{AdS},$ respectively ($Y_P=\eta_{PQ}Y^Q$).
$\Lambda$ and $\tilde{\Lambda}$ are the Lagrange multipliers imposing the two
hypersurface conditions $Y_P Y^P = -1$ and $X_M X_M =1.$
The action (\ref{1}) is to be supplemented with the conformal gauge constraints
\begin{equation}\label{3}
	\dot{Y}_P \dot{Y}^P+Y'_P Y'^P+\dot{X}_M \dot{X}_M+ X'_M X'_M =0, \qquad
	\dot{Y}_P Y'^P+\dot{X}_M X'_M=0
\end{equation}
and the closed-string periodicity conditions
\begin{equation}\label{6}
	Y_P(\tau,\sigma+2\pi)=Y_P(\tau,\sigma), \qquad X_M(\tau,\sigma+2\pi)=X_M(\tau,\sigma).
\end{equation}
The classical equations of motion following from (\ref{1}) are
\begin{equation}\label{2}
	\begin{array}{lll}
		\partial^a \partial_a Y_P-\tilde{\Lambda} Y_P=0,	& \tilde{\Lambda}=\partial^a Y_P \partial_a Y^P,  & Y_P Y^P=-1, \\
		\partial^a \partial_a X_M+\tilde{\Lambda} X_M=0,	& {\Lambda}=\partial^a X_M \partial_a X_M,	& X_M X_M=1.
	\end{array}
\end{equation}
The action is invariant under the $SO(2,4)$ and $SO(6)$ rotations with correspondent conserved
(on-shell) charges
\be\la{char}
	S_{PQ}= \sqrt{\lambda} \int\limits^{2\pi}_0 \frac{d\s}{2\pi}
	(Y_P \dot{Y_Q}-Y_Q \dot{Y_P}),
	\qquad
	J_{MN}= \sqrt{\lambda} \int\limits^{2\pi}_0 \frac{d\s}{2\pi}
	(X_M \dot{X_N}-X_N \dot{X_M})
	\ .
\ee
We are interested in finding ``spinning'' string solutions
that have nonzero values of these charges.

\vspace{20pt}

%\subsection{Coordinates}

It is useful to solve the constraints
\be
	Y_P Y^P = -1 \qquad X_M X_M = 1
\ee
by choosing an explicit parametrization of the embedding
coordinates $Y_P$ and $X_M,$ for example
\begin{equation}	\label{an1}
\ba{c}
\size
	Y_{05}= Y_0+iY_5 =\cosh{\r} e^{i t} \ ,
	\\[8pt]
\size
	Y_{12}= Y_1+iY_2 =\sinh{\r} \cos \te e^{i \phi_1}\ ,  \qquad
	Y_{34}= Y_3+iY_4 =\sinh{\r} \sin \te e^{i \phi_2} \ ;
\ea
\end{equation}
\begin{equation}	\label{X1}
\ba{c}
\size
	X_{12}=X_1+iX_2 = \sin{\g} \cos \psi e^{i \varphi_1} \ , \qquad
	X_{34}=X_3+iX_4 = \sin{\g} \sin \psi e^{i \varphi_2} \ ,
	\\[8pt]
\size
	X_{56}=X_5+iX_6 = \cos{\g} e^{i \varphi_3} \ .
\ea
\end{equation}
Then the corresponding metrics take the form
 \begin{equation}\label{io}
 ds^2_{AdS_5} = - \cosh^2 \rho \ dt^2 + d\rho^2	 + \sinh^2 \rho \ (d \theta^2 + \cos^2 \theta \ d \phi_1^2 +
 \sin^2 \theta \ d \phi_2^2)  \
 \end{equation}
  \begin{equation}\label{io2}
 ds^2_{S^5} = \cos^2 \g \ d\varphi_1^2 + d\g^2 + \sin^2 \g \ (d \psi^2 + \cos^2 \psi \ d \varphi_1^2 +
 \sin^2 \psi \ d \varphi_2^2) . \
 \end{equation}
The Cartan generators
of $SO(2,4)$ corresponding to the three linear isometries of the
\AdS metric
are the translations in the AdS time $t$ and two angles $\phi_a:$
\be \la{EESS}
S_0 \equiv S_{05} \equiv E = \sqrt{\lambda} \E, \quad
S_1 \equiv S_{12} = \sqrt{\lambda} \S_1, \quad
S_2 \equiv S_{34} = \sqrt{\lambda} \S_2 .
\ee
The Cartan generators
of $SO(6)$ corresponding to the three linear isometries of the $S^5$
metric
are the translations in the three angles $\varphi_a:$
\be \la{JJJ}
J_1 \equiv J_{12} = \sqrt{\lambda} \J_1, \quad
J_2 \equiv J_{34} = \sqrt{\lambda} \J_2, \quad
J_3 \equiv J_{56} = \sqrt{\lambda} \J_3.
\ee

\vspace{20pt}

In this paper we also use the other type of embedding
coordinates in \ads:
\begin{equation}
\label{anC}
Y_{05}=y_0 \ e^{i t} \ , \qquad %\\
Y_{12}=y_1 \ e^{i \phi_1} \ , \qquad % \\
Y_{34}=y_2 \ e^{i \phi_2} \ ,
\end{equation}
where
\be y_1= \sinh \r \ \cos \te  \ , \	 \ \
y_2 = \sinh \r \ \sin \te \ \quad {\rm and }
\quad y_0 = \sqrt{1+y_1^2+y_2^2}=\cosh \r \ .
\la{uuu}
\ee
The corresponding \AdS metric takes the form
\be\la{12}
	ds^2_{AdS_5} = - (1+y_1^2 +y_2^2)dt^2
	 -\frac{(y_1 d y_1 +y_2 d y_2)^2}{1+y_1^2 +y_2^2}
	 + d y_1^2 + d y_2^2 + y_1^2 d \phi_1^2
	+ y_2^2 d \phi_2^2 \ .
\ee
Coordinates \rf{an1} we call ``circular'', coordinates \rf{anC} ---
``Cartesian''.

\section{Small rigid strings in \AdS}

\renewcommand{\theequation}{3.\arabic{equation}}
 \setcounter{equation}{0}

Aiming at a better understanding of two-spin solutions in \ads,
in this section we study the case of small strings.

\subsection{Rigid string ansatz}

Our aim here is to study closed strings with two spins,
i.e. rotating in $\phi_{1,2}.$ A natural ansatz for
describing such solutions is the ``rigid'' string ansatz
\cite{ft2} ($0 \leq \s < 2\pi$):
\be\label{jk}
 \ba{c}
	\size t = \kappa \tau \ , \quad	 \phi_1 = \om_1 \tau \ , \quad	\phi_2 = \om_2 \tau
  \quad	 \k, \ \w_1, \ \w_2 = \const \\
	\size y_1 = y_1(\s) \ , \quad y_2 = y_2 (\s) \qquad {\rm or} \qquad
  \size \r= \r(\s)\ , \quad	 \te=\te (\s) \ .% \label{jk}
\ea
\ee
In the ``circular'' coordinates, the string equations of motion and the conformal constraint
for this ansatz read
\be
&& (\theta' \sinh^2 \rho)'=(\omega_1^2-\omega_2^2) \sin \theta \cos \theta\ \sinh^2 \rho \la{ll} \\
&& \rho''- \cosh \rho\ \sinh \rho\ (\kappa^2 + \theta'^2- \omega_1^2
\cos^2 \theta -\omega_2^2 \sin^2 \theta)=0
\label{sys}\\
&&\rho'^2 - \kappa^2 \cosh^2 \rho +\sinh^2 \rho\ ( \theta'^2 +
\omega_1^2 \ \cos^2 \theta	+ \omega_2^2 \ \sin^2 \theta) =0.
\label{hl}\ee
Note, that these equations are not independent,
for example, \rf{sys} is a linear combination of \rf{ll} and \rf{hl}'s first derivative.

In the ``Cartesian'' coordinates, the
string equations of motion and the conformal constraint read
\be
&&\la{cv1}
\frac{y_1 y_1'' + y_2 y_2''}{1+y_1^2+y_2^2} \ y_1 +
\frac{(y_1 y_2' - y_2 y_1')^2+y_1'^2+y_2'^2}{(1+y_1^2+y_2^2)^2} y_1 - y_1'' + (1+\w_1^2) y_1 = 0
\\
&&\la{cv2}
\frac{y_1 y_1'' + y_2 y_2''}{1+y_1^2+y_2^2} \ y_2 +
\frac{(y_1 y_2' - y_2 y_1')^2+y_1'^2+y_2'^2}{(1+y_1^2+y_2^2)^2} y_1 - y_2'' + (1+\w_2^2) y_2 = 0
\\
&&
( y_2' \ y_1 - y_1' \ y_2  )^2 + ( y_1' )^2 + ( y_2' )^2 =
(1+y_1^2+y_2^2) \ (\k^2 \ (1+y_1^2+y_2^2) - \w_1^2 \ y_1^2 - \w_2^2 \ y_2^2 ).
\ee
We may rewrite this system in a more compact form (with only independent equations present):
\begin{eqnarray}
&& ( y_2' \ y_1 - y_1' \ y_2 )'= (\w_1^2 - \w_2^2) \ y_1 \ y_2 \label{Car1} \\
&& ( y_2' \ y_1 - y_1' \ y_2  )^2 + ( y_1' )^2 + ( y_2' )^2 =
	(1+y_1^2+y_2^2) \ (\k^2 \ (1+y_1^2+y_2^2) - \w_1^2 \ y_1^2 - \w_2^2 \ y_2^2 ),
\label{Car2}
\end{eqnarray}
where \rf{Car1} is the difference between \rf{cv1} and \rf{cv2}.

 A general approach to finding	such rigid string solutions in $AdS_5$ (and $S^5$) was
 developed in \ci{afrt} using the reduction of the conformal-gauge string  sigma model
 to the 1D	 Neumann  integrable  model.\foot{A more general rigid string
 ansatz,	 where	in addition to	$ \r= \r(\s),	\	\te=\te (\s) $ one	has
 $\phi_1 = \om_1 \tau  + \a_1(\s) , \	\phi_2 = \om_2 \tau	  + \a_2(\s) $
 and where the	corresponding 1D system is the Neumann-Rosochatius one, was
 considered in \ci{art}.}
 Starting with the $R^{2,4}$
 embedding coordinates (\ref{anC})
 one finds that the equations for $y_0, \ y_1, \ y_2$
  are those of a harmonic oscillator
 constrained to move on a 2d hyperboloid --- an integrable system with two integrals of motion
 $b_1,b_2$	with
  $b_1+b_2 = \k^2 + \om^2_1 + \om_2^2$.
In general, the solutions are expressed in terms of hyperelliptic functions and thus
are not easy  to analyze.
There are few special cases when they simplify --- when
the hyperelliptic surface degenerates into an elliptic one	 and $y_a(\s)$
can be	expressed  in terms of
the standard elliptic  functions. Two of such  cases,  $\w_1 = \w_2,$
corresponding to $S_1 = S_2$ solution, and its boosted analog  with $\k = \w_2 \neq \w_1$
were studied in \cite{TirTs}.
 The existence of simple but  more general solution with
two	 unequal  spins is	an open question.
Recent
study of $\N = 4$ SYM states dual  to  minimal energy spinning string configuration
with two spins ($S_1, S_2$) with $\frac{S_1}{S_2}$ fixed  using
 ABA \cite{rej} suggests that such simple solution might indeed exist.

Here we study small strings with two-spin solutions in \ads,
starting from the flat-space solutions and using
 perturbation theory in the	curvature of \ads.

\subsection{Flat-space limit}\label{Sec2}

In this section, we review the flat-space limit for closed strings
in \ads. Let us start from the expression for the metric in ``circular''
coordinates
\begin{equation}\label{ioR}
	 ds^2_{AdS_5} = - \cosh^2\left(\frac{\rho}{R}\right) \ dt^2 + d\rho^2
	 + R^2 \ \sinh^2\left(\frac{\rho}{R}\right) \
	 (d \theta^2 + \cos^2 \theta \ d \phi_1^2 + \sin^2 \theta \ d \phi_2^2)	 \
	 .
\end{equation}
Here $R$ is the radius of curvature of \ads.

If the size of the string is small $\r=\e \td \r \ll R, \ \e \ll 1,$ % $max \{ \r \} =\r_0 \ll R,$
one
can perform an expansion ($R=1$):% in $\r_0/R:$
\begin{eqnarray}\label{dsexp}
	 ds^2_{AdS_5} = \e^2( - \ d \td t^2 + d \td \rho^2 + \td \r^2 \ d \W_3)
	 + \e^4 \td \r^2 \ \left( - d \td t^2 + \frac{1}{3} \td \r^2 d \W_3 \right)
	 + O\left( {\e^6} \right),
\end{eqnarray}
where $t = \e \td t,$ $d\W_3=d \theta^2 + \cos^2 \theta \ d \phi_1^2 + \sin^2 \theta \ d \phi_2^2.$
The leading term represents the metric of flat $R^{1,4}$ Minkowski space.

\vspace{20pt}

A similar expansion can be performed in terms of the ``Cartesian'' coordinates.
In the limit of small strings
%Let us use scaling $\e $
\be\la{9}
y_1 = \e \td y_1, \qquad y_2 = \e \td y_2, \qquad \e \ll 1,
\ee
where $\e$ defines the size of the string with respect
to the radius of curvature, we have
%Substitution of \rf{9} into (\ref{12}) gives
\be\la{dsexpC}
	ds^2_{AdS_5} = \e^2 \left( -d \td t^2 + d \td y_1^2 +
	d \td y_2^2 + \td y_1^2 d \phi_1^2 + \td y_2^2 d \phi_2^2
	 \right)
	 - \e^4 \left( (\td y_1^2 + \td y_2^2) d \td t^2
	 + \td y_1 d \td y_1 + \td y_2 d \td y_2 \right)+ O(\e^4) \ ,
\ee
where $d t = \e d \td t .$ Again, the leading term
is the metric of flat $R^{1,4}$ Minkowski space.

In this paper we will mainly work with the expansion \rf{dsexpC}.

\vspace{20pt}

In the flat-space limit the string equations of motion and conformal constraint
for the ansatz \rf{jk} become
\begin{equation}\label{7}
\ba{c}
  \td y_1''+ \w_1^2 \td y_1 =0	\qquad
  \td y_2''+ \w_1^2 \td y_2 =0	 \\
   ( \td y_1' )^2 + \w_1^2 \td y_1^2 - \td \k_1^2 =0  \qquad  ( \td y_2' )^2 + \w_2^2 \td y_2^2 - \td \k_2^2=0 \\
  \qquad \ \ \td t = \td \k \t, \qquad \td \k^2 = \td \k_1^2 + \td \k_2^2
\ea
\end{equation}
The solutions of these equations are \cite{afrt}
%\cite{afrt}
\begin{equation} \label{taak}
\ba{c}
\size \td t = \td \k \t, \qquad \td \k^2 =	n^2_1 a_1^2 + n^2_2 a_2^2  \\
\size \td y_1 = {y}_{1}^{\rm flat} = \ a_1	  \sin (n_1 \s) \qquad
\size \td y_2 = {y}_{2}^{\rm flat} = \ a_2 \sin [n_2 (\s + \s_0) ]	 \\
\size \w_1=n_1,
\qquad \w_2=n_2 \ea
\end{equation}
where  $n_i$ are integers and	$\s_0$ is a constant phase shift.
The energy and spins 
are given by
$$
 \E= {\k}  , \ \	 \S_i = \frac{n_i a^2_i}{2}, \qquad \rm{i.e.} \qquad
 \E= \sqrt {	 {2} ( n_1 \S_1 + n_2 \S_2 ) }
$$
or, restoring $\l,$
$$
 E= \sqrt{\l}{\k}  , \ \	 S_i = \sqrt{\l} \frac{n_i a^2_i}{2}, \qquad \rm{i.e.} \qquad
 E= \sqrt {	 {2} \sqrt{\l} ( n_1 S_1 + n_2 S_2 ) }.
$$
To get	the states on the leading  Regge trajectory (having minimal
energy for given  values of the spins) one	is to choose
$n_1=n_2=1$.

Note, that in the case $n_1 = n_2$ and $ \ 2 \s_0 \neq \pi n$
there are also non-Cartan components of the spin present. We have not
mentioned them above, as such solutions
 can always be rotated to\foot{
Let us set, for simplicity, $n_1=n_2=n=1$ and rotate (\ref{taak})
by an angle $\b$ ($\b\neq \frac{\pi}{2} m, \ m \in Z$) in $Y_1 Y_3$ and $Y_2 Y_4$
planes
\begin{equation}
\left(
\ba{c}
	a_1 \sin (\s) \\
	a_2 \sin (\s + \s_0)
\ea
\right)
\rightarrow
\left(
\ba{c}
	\sqrt{(a \cos \b - b \sin \b \cos \s_0)^2+(b \sin \b \sin \s_0)^2}	\sin (\s-\vp_1) \\
	\sqrt{(a \sin \b + b \cos \b \cos \s_0)^2+(b \cos \b \sin \s_0)^2} \cos (\s -\vp_2)
\ea
\right),
\end{equation}
where
\be
&& \sin \vp_1 = \frac{b \sin \b \sin \s_0}{\sqrt{(a \cos \b - b \sin \b \cos \s_0)^2+(b \sin \b \sin \s_0)^2}}\\
&& \sin \vp_2 = \frac{a \sin \b + b \cos \b \cos \s_0}{\sqrt{(a \sin \b + b \cos \b \cos \s_0)^2+(b \cos \b \sin \s_0)^2}}
\ee
When $\vp_1 = \vp_2=\vp_0$ or equivalently
$
	\tan 2\b = \frac{a}{b \cos \s_0},
$
the rotated solution is indeed of the form (\ref{taak2}), with $\s \to \s - \vp_0.$
}
\begin{equation}
\label{taak2} \ba{c}
\size {y}_{1}^{\rm flat}  = a	\sin (n \s), \qquad \size {y}_{2}^{\rm flat}  = b \cos (n \s )	\\
\size \w_1=\w_2=n, \qquad \k^2 = n^2 (a^2 + b^2) \ea
\end{equation}
i.e. ones without non-Cartan components.

\subsection{Curvature corrections to the flat-space solutions in \AdS}\label{Sec3}

Expansions \rf{dsexp} and \rf{dsexpC} suggest the possibility
that solutions in full \AdS
may be constructed as 
\begin{equation}\la{10}
	\ba{ll}
	\size
	y_1(\s) = \e \ y_1^{flat}+ \e^3 z_1(\s)+ \e^5 z_3(\s)+...\\
	\size
	y_2(\s)= \e \ y_2^{flat}+ \e^3 z_2(\s)+ \e^5 z_4 (\s)+...\\
	\ea
\end{equation}
where the first term corresponds to the flat-space solution
\rf{taak}, while the others may be found by using perturbation theory in the curvature of \ads.

Here we will be interested in the first subleading corrections only.

Let us look for a solution of (\ref{Car1}), (\ref{Car2}) in the form:
\begin{equation}\label{n1n2}
	\ba{ll}
	\size
	y_1(\s) = \e \ a \sin(n_1 \s)+ \e^3 z_1(\s)\\
	\size
	y_2(\s)= \e \ b \sin(n_2 (\s+\s_0))+ \e^3 z_2(\s),\\
	\ea
\end{equation}
where $n_{1,2} \in Z$ and
\be\label{wwk}
\ba{ll}
	\size
	\w_1= n_1 (1 + \e^2 \tilde{\w}_1), \qquad
	\size
	\w_2= n_2( 1 + \e^2 \tilde{\w}_2)  \\
	\size
	\k=\e \ \k_0 + \e^3 \ \k_1, \qquad \k_0^2 = a^2 n_1^2+b^2 n_2^2.
\ea
\ee
Here $\td \w_i,$ and $\k_1$ are curvature corrections to $\w_i$ and $\k,$ respectively.

From (\ref{Car1}), (\ref{Car2}) one obtains the following system of equations:
\begin{equation}\label{eq21}
	\ba{ll}
		\size -b \ {\sin}((\s+\s_0) n_2) \ (n_1^2 z_1+z_1'')+ a \sin(\s n_1) \left(n_2^2
		z_2+z_2''\right)\\ [10pt]
		\qquad \qquad \qquad \qquad \qquad \qquad \size =  2 a b \ {\sin}((\s+\s_0) n_2) \sin(\s n_1) (n^2_1 \tilde{\w}_1-n^2_2
		\tilde{\w}_2)
\ea
\end{equation}
\begin{equation}\label{eq22}
	\ba{ll}
		\size 2 a \ (\cos(n_1 \s) n_1 z_1' + \sin(n_1 \s) n_1^2
		z_1)
		+2 b \ (\cos (n_2(\s+\s_0)) n_2 z_2'+\sin(n_2 (\s+\s_0)) n_2^2 z_2)\\[5pt]
		\qquad \size = 2 \chi + \frac{1}{4} (a^2 n_1 \sin(2 n_1 \s) + b^2 n_2 \sin(2 n_2 (\s+\s_0))	 )^2
		\\[5pt]
		\size 
		+ a^2 \ \sin^2(n_1 \s) \left[ a^2 n_1^2+b^2 n_2^2 - 2 a^2 n^2_1 \tilde{\w}_1 \right]
		+ b^2 \ \sin^2(n_1 (\s+\s_0)) \left[ a^2 n_1^2+b^2 n_2^2 - 2 b^2 n^2_2 \tilde{\w}_2
		\right]. 
	\ea
\end{equation}
Here $\chi^2=\k_1^2 \k_0^2.$ The equations for $z_1$ and $z_2$
may be separated in the following way. Differentiate both sides of (\ref{eq22}).
The left-hand side reads
\begin{equation}\label{44}
\ba{ll}
	\size \left( a \ (\cos(n_1 \s) n_1 z_1' + \sin(n_1 \s) n_1^2 z_1) + b \ (\cos (n_2(\s+\s_0)) n_2 z_2'+\sin(n_2 (\s+\s_0)) n_2^2 z_2)
	\right)'\\[5pt]
	   \qquad \qquad \size =  a \cos(n_1 \s) n_1 \left(n_1^2 z_1+z_1''\right)+ b \cos(n_2(\s+\s_0)) n_2 \left(n_2^2 z_2[s]+\left(z_2\right)''[s]\right).
\ea
\end{equation}
Then, compare \rf{44}
with the left-hand side of \rf{eq21}. After some rearrangements we obtain
\be
&&
z_1'' + n_1^2 z_1 = 2 a \sin(n_1 \s) \left[ a^2 n_1^2 \cos^2 (n_1 \s)+b^2 n_2 \cos^2 (n_2(\s+\s_0))-n^2_1 \tilde{\w}_1 \right]
\label{45} \\
&&
	z_2'' + n_2^2 z_2 = 2 b \sin(n_2(\s+\s_0)) \left[ a^2 n_1^2 \cos^2 (n_1 \s)+b^2 n_2 \cos^2 (n_2(\s+\s_0))-n^2_2 \tilde{\w}_2
	\right].
\label{46}
\ee
These equations can be readily solved:
\begin{itemize}
\item
If $n_1 = n_2=n$ one finds
\begin{equation}\label{s401}
	\ba{ll}
\size z_1 = C_1 \sin(n \s)+C_2 \cos(n \s)
\\[6pt]
\size \qquad -\frac{ 1}{4} a \ n \s \left[(a^2+2 b^2 - 4
\tilde{\w}_1) \cos(n \s)-b^2 \cos(n \s+2 n \s_0)\right]
\\[6pt]
\size \qquad -\frac{a}{16} \left[a^2 \sin(3 n \s)+b^2 \sin(3 n \s+2
n \s_0) +2 b^2 \sin(n\s+2 n \s_0) \right.
\\[6pt]
\qquad \qquad \qquad \qquad \size \left.- 2 \sin(n \s) \left( a^2+2
b^2-4 \tilde{\w}_1\right)\right]
\\[6pt]
\size z_2 = C_3 \cos(n \s)+C_4 \sin(n \s)
\\[6pt]
\size \qquad -\frac{1}{4} b \ n \s \left[(b^2 + 2 a^2  - 4
\tilde{\w}_2) \cos(n \s+ n \s_0) - a^2 \cos(n \s-n \s_0) \right]
\\[6pt]
\size \qquad
 -\frac{b}{16} \left[b^2 \sin(3 n \s+ 3 n \s_0)	 +
 a^2 \sin(3 n \s+n \s_0)+ 2 a^2 \sin(n \s- n \s_0)\right.
 \\[6pt]
\qquad \qquad \qquad \qquad \size \left. -2 \sin(n \s+ n \s_0)
\left(b^2 + 2 a^2-4 \tilde{\w}_2\right)\right].
	\ea
\end{equation}
Here $C_i \ (i = 1,2,3,4)$ are integration constants.

The closed-string periodicity condition \rf{6}
requires $z_1, \ z_2$ being periodic in $\s,$ i.e. the linear terms
must vanish:
\begin{equation}
	\ba{ll}\label{per}
\size (a^2+2 b^2 - 4 \tilde{\w}_1)
\cos(n \s)-b^2 \cos(n \s+2 n \s_0) =0 \\[6pt]
\size (b^2 + 2 a^2	- 4 \tilde{\w}_2) \cos(n \s+ n \s_0) - a^2
\cos(n \s-n \s_0)=0	.   \ea
\end{equation}
These equations can be solved for constant values of $\td \w_1, \td \w_2$ only
\bei
\item
in the elliptic string case,
when $2 \s_0 n=\pi	 + 2\pi m, $ $m \in Z,$
\be \la{15a}
\tilde{\w}_1 = \frac{1}{4} (a^2+3 b^2), \qquad \tilde{\w}_2 =
\frac{1}{4} (3 a^2+b^2).
\ee
This case is considered in section \ref{Sec4}.
\item
in the folded string case, when $2 \s_0 n= 2\pi m, $ $m \in {Z}$
\be \la{15b}
 \tilde{\w}_1 = \tilde{\w}_2 = \frac{1}{4}(a^2+b^2).
\ee
This case is considered in section \ref{Sec5}.
\eei
The restriction on $\s_0$ might first look surprising.
One can always rotate (\ref{taak}) with arbitrary $\s_0$ to (\ref{taak2})
(see section \ref{Sec2}) and, using the method given above, find
the curvature corrections to any flat-space solution with $n_1=n_2.$
However, rotating back, we would not remain
in the framework of the rigid string ansatz as the
frequencies $\w_1$ and $\w_2$ are now different (see \rf{15a}).
\item
If $n_1 \neq n_2$ one finds
\begin{equation}\la{s402}
	\ba{ll}
\size z_1 = C_1 \sin(n_1 \s)+C_2 \cos(n_1 \s)
\\[6pt]
\size \qquad +\frac{a b^2}{4 \left(n_1^2-n_2^2\right)} \ \left[-
\cos(n_1 \s) \sin(2 n_2 (\s+\s_0)) n_1 n_2+ \cos(2 n_2 (\s+\s_0))
\sin(n_1 \s) n_2^2\right]
\\[6pt]
\size \qquad  -\frac{a^3}{16 } \left[ \sin(3 n_1 \s) - 2 \sin(n_1
\s)\frac{a^2 n_1^2 + 2 b^2 n_2^2 - 4 n_1^2 \tilde{\w}_1}{n_1^2}
\right.
\\[6pt]
\left. \size \qquad \qquad \qquad \qquad \qquad \qquad \qquad \qquad
 + 4 n_1 \s
\cos(n_1 \s) \frac{a^2 n_1^2 + 2 b^2 n_2^2 - 4 n_1^2
\tilde{\w}_1}{n_1^2} \right]
\\
\size z_2 = C_3 \cos( n_2(\s+\s_0))+C_4 \sin( n_2(\s+\s_0))
\\[6pt]
\size \qquad -\frac{b a^2}{4 \left(n_1^2 - n_2^2\right)} \ \left( -
\cos( n_2(\s+\s_0)) \sin(2 n_1 \s) n_1 n_2+ \cos(2 n_1 \s) \sin(n_2
(\s+\s_0)) n_1^2\right)
\\[6pt]
\size \qquad -\frac{b^3}{16 } \left[ \sin(3 n_2 (\s+\s_0)) - 2
\sin(n_2 (\s+\s_0))\frac{b^2 n_2^2 + 2 a^2 n_1^2 - 4 n_2^2
\tilde{\w}_2}{n_2^2} \right.
\\[6pt]
\left. \size \qquad \qquad \qquad \qquad \qquad \qquad \qquad \qquad
 - 4 n_2 \s
\cos(n_2 (\s + \s_0)) \frac{b^2 n_2^2 + 2 a^2 n_1^2 - 4 n_2^2
\tilde{\w}_2}{n_2^2} \right] . \ea
\end{equation}
Here $C_i \ (i = 1,2,3,4)$ are integration constants.

The closed-string periodicity condition \rf{6} requires
the linear terms vanish:
\be
	a^2 n_1^2 + 2 b^2 n_2^2 - 4 n_1^2 \tilde{\w}_1 = 0, \qquad
	b^2 n_2^2 + 2 a^2 n_1^2 - 4 n_2^2 \tilde{\w}_2 = 0.
\ee
Then (for any value of $\s_0$) we have
\begin{equation}\la{15c}
	\ba{ll}
\size \tilde{\w}_1 = \frac{a^2 n_1^2+2 b^2 n_2^2}{4 n_1^2} \ , \qquad
\size \tilde{\w}_2 = \frac{2 a^2 n_1^2+b^2 n_2^2}{4 n_2^2} \ . \ea
\end{equation}
This case is considered in section \ref{Sec6}.
\end{itemize}

In fact, the restriction on $\s_0$ in the $n_1 = n_2$ case
singles out the solutions with zero non-Cartan components of
spin. Indeed, for $n_1 \neq n_2$ there are no such components for
any value of $\s_0,$ while for $n_1 = n_2$ they vanish only if
$2 \s_0 = \pi m.$

Only flat-space solutions with zero non-Cartan components
of spin receive curvature corrections in the framework of the
rigid string ansatz. An attempt to find the corrections to
the solutions with nonvanishing non-Cartan components leads out
of the rigid string ansatz.

\subsection{The elliptic string solution ($n_1=n_2$)}\label{Sec4}

Curvature corrections to the string solution with
 $n_1=n_2=n$ and $2 \s_0 n=\pi	+ 2\pi m, $ $m \in Z$
are (see \rf{s401}, \rf{15a})
\begin{equation}\label{zz1}
	\ba{ll}
\size z_1 = C_1 \sin(n \s)+C_2 \cos(n \s)
 -\frac{1}{16} a \ (a^2 - b^2) \sin(3 n \s)
\\[6pt]
\size z_2 = C_3 \cos(n \s)+C_4 \sin(n \s)
 +\frac{1}{16} b \ (b^2-a^2) \cos(3 n \s).
	\ea
\end{equation}
Here vanishing of non-Cartan components of spin requires
$aC_4 = - b C_2.$

Recall that in order to get from the system (\ref{eq21}), (\ref{eq22})
to (\ref{45}), (\ref{46}), we take a derivative from
(\ref{eq22}), thus we must check if it is satisfied. %$C_i$ should be constrained.
Substituting \rf{zz1} into (\ref{eq22}), one finds% gives
\begin{equation}\label{chi1}
	-16 \chi -3 n^2 \left(a^2-b^2\right)^2+16 n^2 \left(a C_1+b
	C_3\right) = 0 \ .
\end{equation}
Then the classical energy of the string reads
\begin{equation}\label{En}
	\E_{n_1 = n_2}=\sqrt{2 n (\S_1+ \S_2)}\left[1+\frac{3}{8 n} \ ( \S_1 + \S_2 )+\frac{1}{2 n}\frac{\S_1 \S_2}
	{\S_1 + \S_2}+O(\S_i\S_j) \right]
\end{equation}
or, restoring $\l,$
\begin{equation}\label{Ena}
	E_{n_1 = n_2}=\sqrt{2 n \sqrt{\l} (S_1+ S_2)}\left[1+
	\frac{3}{8 n \sqrt{\l}} \ ( S_1 + S_2 )+\frac{1}{2 n \sqrt{\l}}\frac{S_1 S_2}
	{S_1 + S_2}+O(\l^{-1}) \right] \ .
\end{equation}
This expression is a generalization of circular and folded string
cases (for a review see Appendix \ref{AppA} and references therein).
In the limit $\S_1=\S, \ \S_2 = 0,$ it gives the small-spin
expansion of the classical energy of the folded string (see \rf{A14})
\begin{equation}
	\E=\sqrt{2 n \S}\left(1+\frac{3 \S}{8 n} +O(\S^2) \right);
\end{equation}
in the limit $\S_1 = \S_2 = \S$ --- the small-spin
expansion of the classical energy of the circular string (see \rf{A5})
\begin{equation}
	\E=2\sqrt{n \S}\left(1+\frac{\S}{n} + O(\S^2) \right) .
\end{equation}

\subsection{The folded string solution ($n_1=n_2$)}\la{Sec5}

Curvature corrections to the string solution with
$n_1=n_2=n$ and $2 \s_0 n=\pi + 2\pi m, $ $m \in Z$ are
(see \rf{s401}, \rf{15b})
\begin{equation}\label{zz2}
	\ba{ll}
\size z_1 = C_1 \sin(n \s)+C_2 \cos(n \s)
 -\frac{1}{16} a \ (a^2 + b^2) \sin(3 n \s)
\\[6pt]
\size z_2 = C_3 \cos(n \s)+C_4 \sin(n \s)
 -\frac{1}{16} b \ (b^2+a^2) \cos(3 n \s).
	\ea
\end{equation}
Here vanishing of non-Cartan components of spin requires
$aC_4 = - b C_2.$

From (\ref{eq22}), one obtains the constraint on $C_i:$
\begin{equation}\label{chi2}
	-16 \chi - 3 n^2 \left(a^2+b^2\right)^2+16 n^2 \left(a C_1+b
	C_3\right) = 0.
\end{equation}
Then the classical energy of the string reads
\begin{equation}
	\E=\sqrt{2 n S} \left(1+\frac{3}{8 n}\S+O(\S^2) \right)
	\quad {\rm or} \quad
	E=\sqrt{2 n \sqrt{\l} S} \left(1+\frac{3}{8 n \sqrt{\l}} \ S+O(\l^{-1}) \right),
\end{equation}
i.e. coincides with the small-spin expansion of the classical
energy of the folded string \rf{A14}.

\subsection{$n_1 \neq n_2$ solutions }\la{Sec6}
Curvature corrections to the string solution with $n_1 \neq n_2$
and arbitrary phase shift $\s_0$ are (see \rf{s402}, \rf{15c})
\begin{equation}\label{zz3}
	\ba{ll}
\size z_1 = C_1 \sin(n_1 \s)+C_2 \cos(n_1 \s)
\\[6pt]
\size \qquad +\frac{a b^2}{4 \left(n_1^2-n_2^2\right)} \ \left[-
\cos(n_1 \s) \sin(2 n_2 (\s+\s_0)) n_1 n_2+ \cos(2 n_2 (\s+\s_0))
\sin(n_1 \s) n_2^2\right]
\\[6pt]
\size \qquad \qquad \qquad \qquad -\frac{a^3}{16 } \sin(3 n_1 \s) 
\\[10pt]
\size z_2 = C_3 \sin( n_2(\s+\s_0)) + C_4 \cos( n_2(\s+\s_0))
\\[6pt]
\size \qquad +\frac{b a^2}{4 \left(n_2^2 - n_1^2\right)} \ \left( -
\cos( n_2(\s+\s_0)) \sin(2 n_1 \s) n_1 n_2+ \cos(2 n_1 \s) \sin(n_2
(\s+\s_0)) n_1^2\right)
\\[6pt]
\size \qquad \qquad \qquad \qquad -\frac{b^3}{16 } \sin(3 n_2
(\s+\s_0)). \ea
\end{equation}
From (\ref{eq22}), one obtains the constraint on $C_i:$
\begin{equation}\label{chi3}
	-16 \chi -3\left(a^4 n_1^2+ b^4 n_2^2\right) + 16 \left( a \ n_1^2 \ C_1 + b \ n_2^2 \ C_3\right)=0.
\end{equation}
Then the classical energy of the string %from \rf{zz3} and \rf{chi3}
reads
\begin{equation}\label{En2}
	\E_{n_1 \neq n_2}=\sqrt{2 n_1 \S_1+2 n_2 \S_2}\left[  1+\frac{3}{8} \ \frac{(\S_1+ \S_2)^{2}}{n_1 \S_1+ n_2 \S_2}
	+\frac{1}{2} \ \frac{\S_1 \S_2}{n_1 \S_1+n_2 \S_2} \left( \frac{n_1}{n_2} + \frac{n_2}{n_1} - \frac{3}{2}
	\right)+O(\S_i \S_j) \right]
\end{equation}
or, restoring $\l,$
\begin{equation}\la{En2a}
\ba{ll}
\size
	 E_{n_1 \neq n_2}=\sqrt{2 \sqrt{\l} ( n_1 \S_1 + n_2 \S_2)}\left[	1+\frac{3}{8 \sqrt{\l} } \
	 \frac{(\S_1+ \S_2)^{2}}{n_1   \S_1+ n_2 \S_2}\right.
	 \\
\size
	\qquad \qquad \qquad \qquad \qquad
	 \left.+\frac{1}{2 \sqrt{\l}} \ \frac{\S_1 \S_2}{n_1 \S_1+n_2 \S_2} \left( \frac{n_1}{n_2} + \frac{n_2}{n_1} - \frac{3}{2}
	 \right)
	 + O(\l^{-1})\right] .
\ea
\end{equation}
Note, that in the limit $n_1=n_2=n$ expression \rf{En2a}
becomes
$$
	E_{n_1 \neq n_2} \stackrel{n_1=n_2}{\longrightarrow}  \sqrt{2 \sqrt{\l} n ( \S_1 + \S_2)}\left[	 1+\frac{3}{8 n \sqrt{\l} } \ \frac{(\S_1+ \S_2)^{2}}{\S_1+ \S_2}
	+\frac{1}{2 n \sqrt{\l}} \ \frac{\S_1 \S_2}{\S_1+\S_2} \left(  \frac{1}{2}
	\right)+ O(\l^{-1})\right],
$$
which differs from \rf{Ena} by the factor of $1/2$ in the third term in the brackets.

This discontinuity
may indicate that there are deep differences between solutions
 with $n_1 \neq n_2$ and more symmetrical ones with $n_1 = n_2.$

%%%%%%%%%%%%%%%%%%%%%%%%%%%%%%%%%%%%%%%%%%%%
\section{Small-string limit of the exact string solutions in \AdS}\label{w1w2}
%%%%%%%%%%%%%%%%%%%%%%%%%%%%%%%%%%%%%%%%%%%%
\renewcommand{\theequation}{4.\arabic{equation}}
 \setcounter{equation}{0}

In this section we investigate the connection between
small- (flat-space) and large-spin limits of
two-spin string solutions in \ads.
In the particular cases	 of $\w_1 = \w_2$ and $\k=\w_2 \neq \w_1$
the general solutions in \AdS were found in \ci{TirTs}.
It was discussed there,
for $\k=\w_2 \neq \w_1$ case, that strings which admit large-spin limit do
not have the small-spin one and vice-versa.
Thus we study only solutions with $\w_1 = \w_2,$ corresponding to $S_1 = S_2$ case.

When $\om_1 = \om_2=\om,$ string sigma model equations reduce to
\begin{eqnarray}
&& \theta'= \frac{c}{\sinh^2 \rho}	 \label{sqr} \\
&& \rho'^2=\kappa^2 \cosh^2 \rho - \frac{c^2}{\sinh^2 \rho}-\omega^2
\sinh^2 \rho  \ , \label{con}
\end{eqnarray}
where $c$ is an integration constant.
The solution for $\r$ is \cite{TirTs}
\begin{equation}\label{sol1}
\cosh \rho= \frac{\sqrt{a_{-}}}{\dn[ \sqrt{a_+ (\om
^2-\kappa^2)}\sigma, \frac{a_+ - a_-}{a_+}]} \ .
\end{equation}
Here
\begin{equation}\label{a+-}
a_{\pm}= \frac{2 \omega^2 - \kappa^2 \pm \sqrt{\kappa^4-4 c^2
(\omega^2-\kappa^2)}}{2 (\omega^2- \kappa^2)}
\end{equation}
define the size of the string: $\sqrt{a_-} \leq
\cosh \rho \leq \sqrt{a_+}.$
Parameters $\k, \ \om, \ c$ are related to $a_\pm$ as
\begin{equation}
c^2=(a_{+}-1)(a_{-}-1) (\omega^2 -\kappa^2), \quad \quad \kappa^2=
\omega^2 \frac{a_{+}+a_{-}-2}{a_{+}+a_{-}-1} \ . \label{opii}
\end{equation}
Solution \rf{sol1} is valid for $\sqrt{a_-} \leq
\cosh \rho \leq \sqrt{a_+}$ only.

\vspace{20pt}

Let us expand \rf{sol1} in the small-string limit. 

When the size of the string is small
with respect to the curvature of \AdS space ($R=1$), one has
\begin{eqnarray}\la{sm}
	\ba{c}
	\size
	a_+ = \cosh \r_{max} = 1 + \e^2 \ a^2 + \e^4 \ A +O(\e^6) \	 \\
	\size
	 a_- = \cosh \r_{min} = 1 + \e^2 \ b^2 + \e^4 \ B +O(\e^6)\
	\ea
	 \qquad \e \ll 1 \ .
\end{eqnarray}
In what follows we omit orders higher than $\e^4.$

 In that limit the elliptic modulus of $\dn$ in \rf{sol1} is small
$$
\frac{a_+ -
a_-}{a_+}=\frac{\e^2 \ (a^2 - b^2) + \e^4 \ (A-B) }{1 + \e^2 \ a^2 +
\e^4 \ A} \sim \e^2 \ll 1 \ ,
$$
so we can perform an expansion
\begin{equation}\label{11}
\ba{ll} \size \cosh \r=1 + \e^2 \ \frac{1}{2} (a^2 \sin^2 (W \s)+b^2 \cos^2 (W \s)) \\[6pt]
\displaystyle \qquad \qquad \ \ + \ \e^4 \ \frac{1}{8} \big[ \ W
(a^4 - b^4) \ \s \ \sin(2 W \s) - \frac{1}{4} \ (a^2 - b^2)^2 \
\sin^2(2 W \s)\big.
\\[10pt]
\displaystyle \qquad \qquad \qquad \ \ \ \ \ \ \ \ \big.- \left(a^4
- 4 A\right) \sin^2(W \s) - \left(b^4 - 4 B\right) \cos^2(W
\s)\big]+ O(\e^6) \ . \ea
\end{equation}
Here $W^2=\om ^2-\kappa^2.$
To satisfy the closed-string periodicity condition, the $\e^2$ and $\e^4$
terms must both be periodic. There are two options:
\bei
\item $W$ is an integer. Then the $\e^2$ term is periodic
and the linearity in the $\e^4$ term
cancels if $a=b.$

\item $W$ has the form $W=W_0 + \e^2 \ W_1.$ Then from \rf{11} we have
\begin{equation}\label{co}
\ba{ll} \size \cosh \r=1 + \e^2 \ \frac{1}{2} (a^2 \sin^2 (W_0 \s)+b^2 \cos^2 (W_0 \s)) \\[6pt]
\displaystyle \qquad \qquad \ \ + \ \e^4 \ \frac{1}{8} \big[ \ (a^2
- b^2) ( 4 W_1 + m (a^2+b^2)	) \ \s \ \sin(2 W_0 \s) -
\frac{1}{4} \ (a^2 - b^2)^2 \ \sin^2(2 W_0 \s)\big.
\\[10pt]
\displaystyle \qquad \qquad \qquad \ \ \ \ \ \ \ \ \big.- \left(a^4
- 4 A\right) \sin^2(W_0 \s) - \left(b^4 - 4 B\right) \cos^2(W_0
\s)\big]. \ea
\end{equation}
The $\e^2$ term is periodic if $W_0$ is an integer and the linearity in the $\e^4$ term
cancels if $a=b$ or
\begin{equation}\label{W}
W_1 = - \e^2 \ \frac{1}{4} \ W_0 \ (a^2+b^2) \ .
\end{equation}
\eei

The case $a=b$ brings us
to the trivial limit of the circular string, so we will not discuss it here.
Let us investigate the other option.

Assuming that $W$ has the form \rf{W}, we get
%Then (\ref{co}) takes the form
\begin{equation}\label{cosh}
\ba{ll} \size \cosh \r=1 + \e^2 \ \frac{1}{2} (a^2 \sin^2 (W_0 \s)+b^2 \cos^2 (W_0 \s)) \\[6pt]
\displaystyle \qquad \qquad \ \ - \ \e^4 \ \frac{1}{8} \left[
\frac{1}{4} \ (a^2 - b^2)^2 \ \sin^2(2 W_0 \s)\right.
\\[10pt]
\displaystyle \qquad \qquad \qquad \ \ \ \ \ \ \ \ \left.+ \left(a^4
- 4 A\right) \sin^2(W_0 \s) + \left(b^4 - 4 B\right) \cos^2(W_0
\s)\right] . \ea
\end{equation}
Making use of \rf{sqr} and \rf{cosh}, one obtains the following
equation for $\th$
\begin{equation}
\ba{ll}
	\size \theta'= \frac{\tilde{c} \ \e^2}{\sinh^2 \rho }
	= \frac{\tilde{c}_0 + \e^2 \ \tilde{c}_1 }{a^2 \sin^2 (W_0 \s)+b^2 \cos^2 (W_0 \s)}
	\\[10pt]
	\size \qquad -	\e^2 \ \tilde{c}_0 \ \frac{2 A \sin^2 (W_0 \s)+2 B \cos^2 (W_0 \s) -
	(a^2 - b^2)^2 \cos^2 (W_0 \s) \sin^2 (W_0 \s)}
	{2 (a^2 \sin^2 (W_0 \s)+b^2 \cos^2 (W_0 \s))^2},
\ea
\end{equation}
where
\be\la{tdc}
c = \tilde{c} \ \e^2 = \e^2 \ \tilde{c}_0 + \e^4 \ \tilde{c}_1 .
\ee
Its solution is
\be
	\theta(\s)= \theta_0(\s) + \e^2 \ \theta_1 (\s),
\ee
where
\be
\ba{ll}
	\size \theta_0(\s) = \frac{\tilde{c}_0}{W_0 \ a b} \arctan\left[ \frac{a}{b} \tan(W_0 \s)
	\right]; \\[8pt]
	\size \theta_1 (\s)
	 =	 \frac{\ \tilde{c}_0}{4 W_0 \ a b}
	\left( a^2 + b^2 - 2 \frac{A}{a^2} - 2 \frac{B}{b^2} + 4 \frac{\tilde{c}_1}{\tilde{c}_0} \right)
	\arctan\left[ \frac{a}{b} \tan(W_0 \s)\right]
	\\[8pt]
	\qquad \size - \frac{\tilde{c}_0}{4 W_0} \left( a^2 - b^2 - 2 \frac{A}{a^2} + 2 \frac{B}{b^2} \right)
	\frac{\cos(W_0 \s) \sin(W_0 \s)}{a^2 \sin^2 (W_0 \s)+b^2 \cos^2 (W_0 \s)}
		-\frac{ \tilde{c}_0}{2 W_0} \arctan\left[ \tan(W_0 \s) \right].
\ea
\ee

This expression, as well as (\ref{sol1}) and \rf{co}, is valid for
$0 \leq W_0 \s \leq \frac{\pi}{2}$ only. Within this interval
$\th$ may change only by a rational value of $\pi:$ $0 \leq \theta \leq
\frac{n}{k}\theta$ and may not gain any
small corrections, otherwise the solution would not satisfy the closed-string
periodicity condition. We must have that
$\theta_1(W_0 \s =0)=\theta_1 (W_0 \s = \frac{\pi}{2})$. The latter gives
the following constraint on $\td c_i, \ A, \ B$
\begin{equation}\label{tildec}
   4 \frac{\tilde{c}_1}{\tilde{c}_0} = 2 \left(\frac{A}{a^2} + \frac{B}{b^2}\right) -(a - b)^2.
\end{equation}

\vspace{20pt}

So far we have not used the relations given in \rf{opii}. Substitution of (\ref{W}) and \rf{tdc} into (\ref{opii}) gives
\begin{equation}
	4 \frac{\tilde{c}_1}{\tilde{c}_0} = 2 \left(\frac{A}{a^2} + \frac{B}{b^2}\right)
	-(a - b)^2+ab.
\end{equation}
Comparing this to (\ref{tildec}), one finds
$$
ab = 0, 
$$
which implies $a=0$ or $b=0$ and brings us to the limit of folded string.

Apart from the
trivial cases of folded and circular strings,
we find that the general
rigid solution with $\w_1 = \w_2$ $(S_1=S_2)$ in \AdS \
admitting the large-spin limit does not have
a  small-spin limit. For more general two-spin
solutions it might still be possible to have both limits.

\section{Chiral solutions for a bosonic string in \RS}

\renewcommand{\theequation}{5.\arabic{equation}}
 \setcounter{equation}{0}

In this section we discuss chiral solutions in \RS.
Such solutions
obey an additional constraint
\begin{equation}
   \partial_+ X_M \partial_- X_M=0,
\end{equation}
where $X_M$ are embedding coordinates
of $R^6$ with the Euclidean metric $\delta_{MN};$ $X_M X_M = 1$
 and
 $$\partial_{\pm}=\frac{\partial}{\partial \sigma_{\pm}}= \frac{1}{2}\left(
\frac{\partial}{\partial \tau} \pm \frac{\partial}{\partial
\sigma} \right),\qquad	\s_\pm = \t \pm \s.$$

We will discuss the string located at the
center of $AdS_5$ and  rotating in $S^5,$ trivially embedded
in $AdS_5$ as $Y_5+iY_0=e^{it},$
 with the global AdS time being $t=\k \tau$ and
$Y_1,...,Y_4=0$ (see (\ref{2})).

The classical string equations in conformal gauge become\footnote{
Chiral solutions may also be considered via Pohlmeyer reduction \cite{GrigTs}.
For example, let only four of $X^M$'s are nonzero.
The reduced model corresponding to the string in $R_t \times S^3$ \cite{lup}
is the complex sine-Gordon (CSG) model
 \begin{equation}
   \td L =
   \del_{+} \a \del_{-} \a	+	\tan^2 { \a }\ \del_{+} \theta \del_{-} \theta
 + \frac{\k^2}{2} \cos 2\a \,.
\end{equation}
The variables $\a$	and $\theta$  are expressed in terms of	 the $SO(4)$ invariant combinations of derivatives	of the original variables  $X_m$ ($m=1,2,3,4$)
\begin{equation}
 \k^ 2 \cos 2\a = \del_{+} X_M \del_{-} X_M \ , \qquad
\k^3 \sin^2 {\a }\ \del_\pm	 \theta
= \mp \frac{1}{2} \epsilon^{MNKL} X_M \del_{+} X_N \del_{-} X_K \del_\pm^2 X_L
\,.
\end{equation} Chiral solutions meet particular case of
 $\a = \frac{\pi}{4}.$
}
\begin{eqnarray}\la{cheq}
   && \displaystyle \partial_- \partial_+ X_M=0 \label{5a} \\
   && \displaystyle \partial_- X_M \partial_+ X_M =0 \label{5b} \\
   && \displaystyle \kappa^2 = 4 \partial_\pm X_M \partial_\pm X_M \label{5c}
\end{eqnarray}

The simplest solution of this kind	is	\cite{art},
$$
\k = 2 a_3 m_3 \ , \quad
 X_1 + i X_2 = a_1 e^{ i m_1 \s_\pm } \ , \quad
 X_3 + i X_4 = a_2 e^{ i m_2 \s_\pm } \ , \quad
 X_5 + i X_6 = a_3 e^{ i m_3 \s_\mp } \ ,
$$
where $\ss a_i = 1$ and $m_i$ are integers. It was recently used
in	\cite{ko}  as a	  model of a quantum string state
with ``small'' quantum numbers.
We expect that more general	 chiral solutions
may also find  useful applications.

Let us consider the ansatz
\begin{equation}\label{chan}
\begin{array}{c}
%	\displaystyle y_0=Y_5+iY_0=e^{i \kappa \tau} \\
   \displaystyle X_{12}=X_1 + i X_2 = a_1 e^{i F_1(\sigma_+) } \\
   \displaystyle X_{34}=X_3 + i X_4 = a_2 e^{i F_2(\sigma_+ )} \\
   \displaystyle X_{56}=X_5 + i X_6 = a_3 e^{i F_3(\sigma_-) }
\end{array}
\end{equation}
where $\ss a_i = 1.$
To satisfy periodicity condition,
$F_1,F_2$ must have the form
\begin{equation}\label{Fi}
F_{i}(\s_+)=m_{i}\s_{+} + \sum\limits_n {f_n^{(i)} \cos(n\s_{+}) + g_n^{(i)} \sin(n\s_{+}) }
\end{equation}
with $f^{i}_n, g^{i}_n$ real and $m_{i}$ integers.

From string equations \rf{5a}, \rf{5b}, \rf{5c} one finds
\begin{eqnarray}
	   && \kappa^2=4 a_1^2 \left( {\partial_{+} F_1} \right)^2+
	   4 a_2^2 \left( {\partial_{+} F_2} \right)^2 , \label{f1f2}
	   \qquad \qquad \k = 2 m_3 a_3,
	   \\
	   && F_3=m_3 \sigma_-.	 \label{f3}
\end{eqnarray}
Let us assume that $F_1$ is an arbitrary function of the form \rf{Fi} and
then $F_2$ is expressed as
\begin{equation}\label{int}
   F_2(\s_+)=\pm \int{\frac{1}{a_2}\sqrt{ a_3^2 m_3^2 -
   a_1^2\left( {\partial_{+} F_1} \right)^2 }}d\s_{+}.
\end{equation}
Being represented as an integral from the periodic function, $F_2$ possess
periodic and linear terms only. So up to adjusting $a_i,$ it has the form \rf{Fi}.

The general solution for the ansatz (\ref{chan}) is
\begin{equation}\label{csol2}
\begin{array}{ll}
   \displaystyle \k = 2 m_3 a_3 \\
   \displaystyle X_{12}= a_1 \, e^{i F_1(\s_{+}) }, \qquad \qquad
   F_{1}(\s_+)=m_{1}\s_{+} + \sum\limits_n {f_n} \cos(n\s_{+}) + g_n \sin(n\s_{+}) \\
   \displaystyle X_{34}= a_2 \, e^{i F_2(\s_{+}) }, \qquad \qquad
   F_{2}(\s_+)=\pm{\frac{1}{a_2} \int d\s_{+} {\sqrt{ a_3^2 m_3^2-
   a_1^2\left({\partial_{+} F_1} \right)^2	  }} }	\\
   \displaystyle X_{56}= a_3 e^{i m_3 \s_{-}}
\end{array}
\end{equation}

In general, the ``phase function $F_2$'' resulting from the integration \rf{int}
is expressed in elliptic functions.
There are few cases when it simplify to elementary ones. Two of them
$F_1(\s_+) = m_1 \s_+, \ F_2(\s_+) = m_2 \s_+$
and
$F_1(\s_+) = \a \cos n\sigma_+ \ , \
F_2(\s_+) = \a \sin n\sigma_+  \ $
are discussed below.

Note that  chiral solutions treat
$\t$ and $\s$  on an equal footing,
i.e.  nontrivial dependence on $\tau$ implies that the
shape of the string is not rigid, in general, so such solutions are similar to ``pulsating'' ones.

%%%%%%%%%%%%%%%%%%%%%%%%%%%%%%%%%%%%%%%%%%%%%%%%%%%%
\subsection{Rigid chiral solutions}\label{lppw}
%%%%%%%%%%%%%%%%%%%%%%%%%%%%%%%%%%%%%%%%%%%%%%%%%%%%

The simplest chiral solution from ansatz \rf{chan} corresponds to
\be
	F_1(\s_+) = m_1 \s_+, \qquad F_2(\s_+) = m_2 \s_+ \ .
\ee
It reads \cite{art}:
\be \label{hhh}
\k = 2 a_3 m_3 \ , \ \ \ \
 X_{12} = a_1 e^{ i m_1 \s_+ } \ , \ \ \ \
 X_{34} = a_2 e^{ i m_2 \s_+ } \ , \ \ \ \
 X_{56} = a_3 e^{ i m_3 \s_- } \ ,
\ee
where
\be
   a_1^2 m_1^2 + a_2^2 m_2^2 = a_3^2 m_3^2
\ee
with $m_i$ integers and $\ss a_i^2=1.$

Comparing that to \rf{jk}, we see that it is a rigid string solution.
In fact, it is the only possible rigid chiral solution from
ansatz	\rf{chan}.

For fixed $m_i$, the energy is given by the standard flat-space
linear Regge relation
\be
\E = \sqrt{ 2 (m_1 \J_1 + m_2 \J_2 + m_3 \J_3)}, \qquad
m_3 \J_3 = m_1 \J_1 + m_2 \J_2 \ ,
\ee
where expressions for spins are
\be\label{ESSplane}
	\J_1 = a_1^2 m_1, \qquad \J_2 = a_2^2 m_2, \qquad \J_3 = a_3^2 m_3 \ .
\ee
Restoring $\l,$ we get
\be
E = \sqrt{ 2 \sqrt{\l} (m_1 J_1 + m_2 J_2 + m_3 J_3)}, \qquad
m_3 J_3 = m_1 J_1 + m_2 J_2 \ .
\ee

Note, that the non-Cartan components are zero only for
$m_1 \neq m_2.$ If $m_1 = m_2$ the solution can always be rotated to
a two-spin one ($\S_2 = 0$).

\subsection{Sine-cosine solutions}

\begin{figure}[ht]
\centerline{\includegraphics[scale=.8]{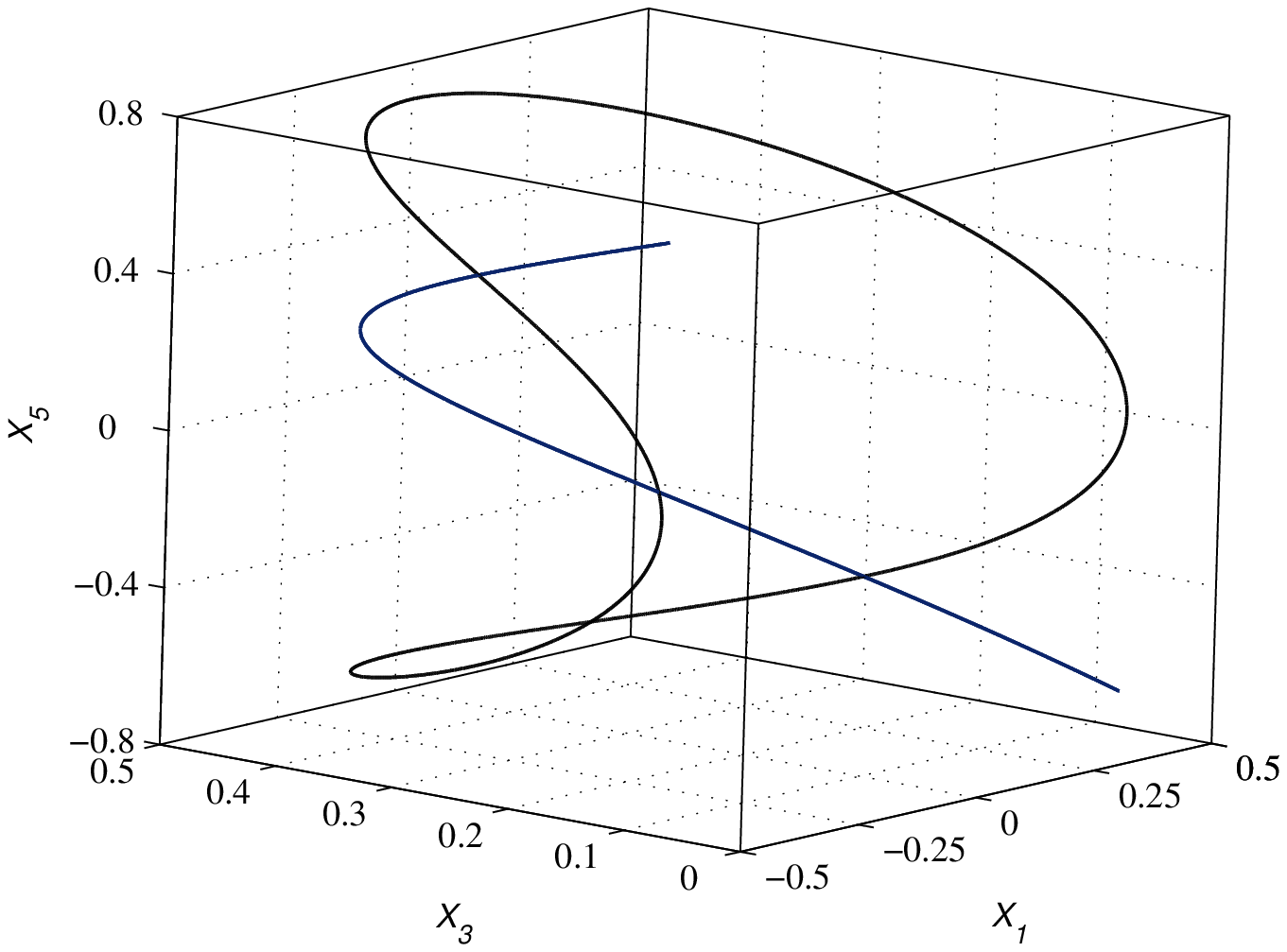}}
\caption{Shape of the string for $n=1, \ m=1, \ \g = \frac{\pi}{4}:$}
{\qquad \qquad \qquad a bended circle at $\t=0$ and folded (in projection
on $X_1X_3X_5$) at $\t=\frac{\pi}{2}$}
\label{StringShape}\nonumber
\end{figure}

A particularly simple nontrivial solution from ansatz \rf{chan} corresponds to
\be\la{sc}
F_1(\s_+) = \a \cos n\sigma_+ \ , \qquad
F_2(\s_+) = \a \sin n\sigma_+ \ . 
%F_3(\s_-) = m \s_- \ .
\ee
It reads
\begin{equation}\label{8a}
\begin{array}{ll}
   \displaystyle \k = 2 m \sin\gamma
\\
   \displaystyle X_1 = \frac{1}{\sqrt{2}}\cos \gamma \
   \sin \left[	   { \sqrt{2} \ \frac{m}{n} \tan{\gamma} \, \cos (n \s_+) } \right]
\\ [10pt]
   \displaystyle X_2 = \frac{1}{\sqrt{2}}\cos \gamma \
   \sin \left[{\sqrt{2} \ \frac{m}{n} \tan{\gamma} \, \sin (n \s_+) }\right]
\\ [10pt]
   \displaystyle X_3 = \frac{1}{\sqrt{2}}\cos \gamma \
   \cos \left[ { \sqrt{2} \ \frac{m}{n} \tan{\gamma} \, \cos (n \s_+) }		\right]
\\ [10pt]
   \displaystyle X_4 = \frac{1}{\sqrt{2}}\cos \gamma \
   \cos \left[{\sqrt{2} \ \frac{m}{n} \tan{\gamma} \, \sin (n \s_+) }\right]
\\ [10pt]
   \displaystyle X_5 = \sin \gamma \cos( m \s_-)
\\
   \displaystyle X_6 = \sin \gamma \sin ( m \s_-)
\end{array}
\end{equation}
Here $n,m$ are integers, %$\gamma \neq \frac{\pi}{2}l,$
$a_1=a_2=\frac{1}{\sqrt{2}}\cos \gamma \neq 0,$
$a_3=\sin \gamma \neq 0.$
Snap shots of the string at $\t=0$ and $\t = \frac{\pi}{4}$
are given in figure \ref{StringShape}. One could see how it changes
shape: a bended circle at $\t=0$ and folded (in projection
on $X_1X_3X_5$) at $\t=\frac{\pi}{2}.$

The energy and spins are (see Appendix
\ref{appC}):
\begin{equation}\label{ESSinCos}
\ba{ll}
   \size \E= 2 m \sin\gamma \  \\
   \size \J_1=\J_{12}=\frac{1}{2} m \sin(2\gamma) \ {\rm BesselJ}_1\left[2 \frac{m}{n} \tan \gamma\right] \ \\
 %	\size \J_2=\J_{34}=0 \	\\
   \size \J_3=\J_{56}= m \sin^2 \gamma \
\end{array}
\end{equation}
For fixed $n$ and $m$ we find
\begin{equation}
\begin{array}{c}
   \displaystyle   \E=2\sqrt{m \J_3}, \qquad% \\
   \displaystyle   \J_1=\sqrt{\J_3 (m-\J_3)} {\rm BesselJ}_1 \left[ 2 \frac{m}{n} \sqrt{ \frac{\J_3}{m-\J_3} }
   \right],
\end{array}
\end{equation}
or, restoring $\l,$
\begin{equation}
\begin{array}{c}
   \displaystyle   E=2\sqrt{ m \sqrt{\l} \ J_3}, \qquad % \\
   \displaystyle   J_1=\sqrt{m \sqrt{\l} \ J_3 \left(1 -\frac{J_3}{m\sqrt{\l}}\right)}
	{\rm BesselJ}_1 \left[ 2 \frac{m}{n} \sqrt{\frac{\frac{J_3}{m\sqrt{\l}}}{1 - \frac{J_3}{m\sqrt{\l}}} }
   \right] .
\end{array}
\end{equation}

\begin{figure}[ht]
\centerline{\includegraphics[scale=.8]{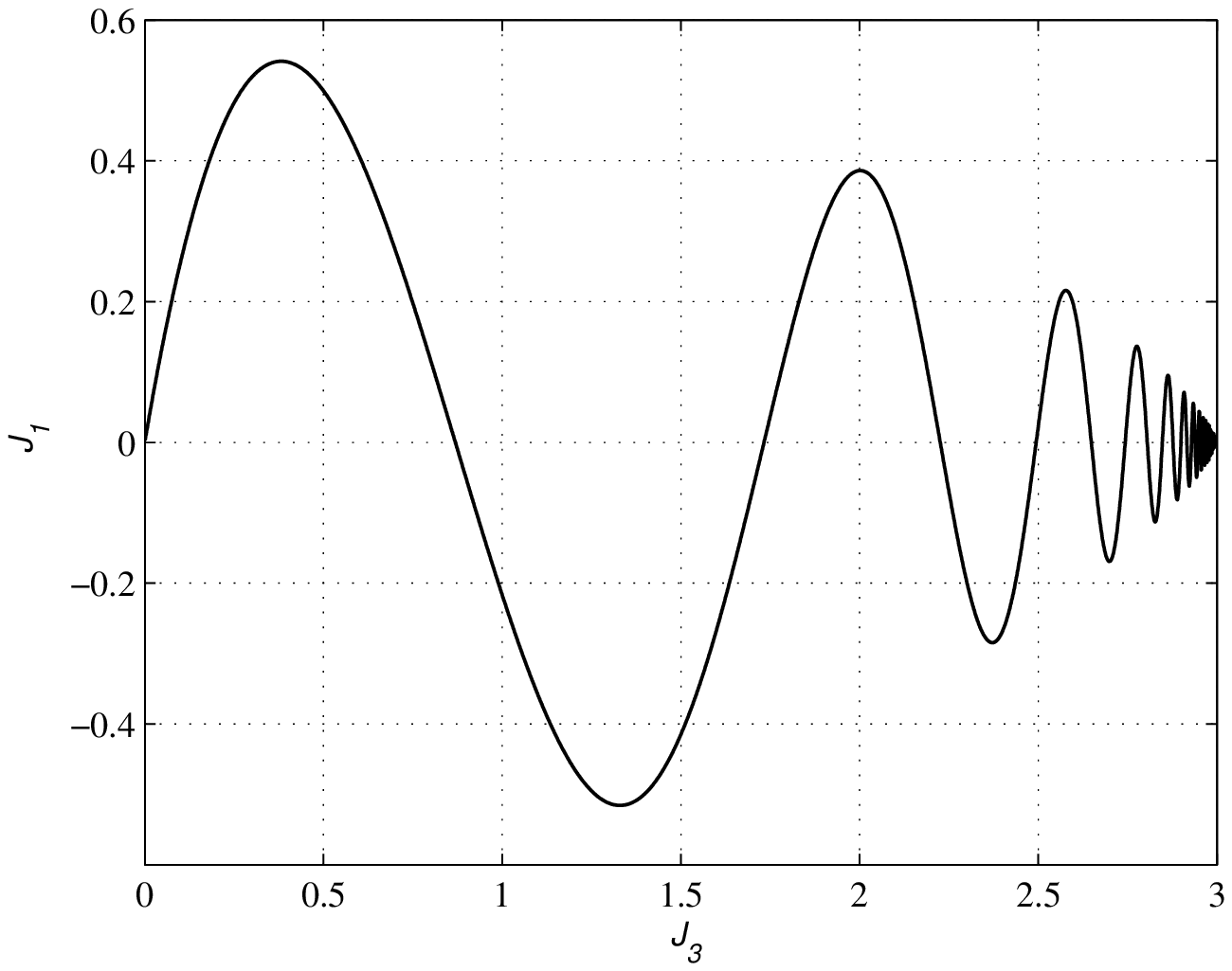}}
\caption{$\J_1(\J_3)$ for $n=1, \ m=3.$
\label{J1J3}}\nonumber
\end{figure}

The dependence $\J_1(\J_3)$ is presented in figure \ref{J1J3}.

Solution \rf{8a} does not admit large-spin limit, as the values of spins are
bounded above:
\be
	J_3 \leq J_3^{max} = m \sqrt{\l}, \qquad J_1 \leq J_1^{max} \approx 0,6 \frac{n \ \sqrt{\l}}{1+\frac{n^2}{m^2}}.
\ee

In the small-spin limit, expanding
the Bessel function in the expression for $\J_1,$ we obtain
\be
\ba{c}
\size
	E=2\sqrt{ m \sqrt{\l} \ J_3}, \qquad % \\
\size
	J_1 = J_3 \frac{m}{n} \left(1 - \frac{1}{2} \frac{m^2}{n^2} \frac{J_3}{m\sqrt{\l}}+O(\l^{-1})\right) .
\ea
\ee

\vspace{20pt}

Let us show that in the limit $n \to 0,$ certain solutions of type \rf{8a} reduce to \rf{hhh}.

In that limit from \rf{8a} one finds
\begin{equation}\la{red}
\begin{array}{c}
   \displaystyle \k = 2 m \sin\gamma
\\
   \displaystyle X_1 = \frac{1}{\sqrt{2}} \cos \gamma, \qquad
   \displaystyle X_3 = \frac{1}{\sqrt{2}} \cos \gamma
\\
   \displaystyle X_2 = \frac{1}{\sqrt{2}}\cos \gamma \
   \sin \left[{\sqrt{2} \ m \tan{\gamma} \s_+ }\right] ,
\qquad
   \displaystyle X_4 = \frac{1}{\sqrt{2}}\cos \gamma \
   \cos \left[{\sqrt{2} \ m \tan{\gamma} \, \s_+ }\right]
\\ [10pt]
   \displaystyle X_5 = \sin \gamma \cos( m \s_-), \qquad
   \displaystyle X_6 = \sin \gamma \sin ( m \s_-)
\end{array}
\end{equation}
Here we omitted the infinite phases in $X_1$ and $X_3$, coming from
$\frac{\cos(n \s_+)}{n},$ as they
do not contribute to the consideration.

Relations \rf{red} look like ones describing rigid
chiral solution with two spins. Indeed one
could rewrite them as
\begin{equation}\la{no}
\begin{array}{c}
   \displaystyle \k = 2 a_3 m \\
\displaystyle X_1 = a, \qquad X_3 = a , \\
\displaystyle X_2 = a \sin ( k \s_+ ), \qquad
   \displaystyle X_4 = a \cos (k \s_+) \\ %[6pt]
   \displaystyle X_5 = c \cos( m \s_- ), \qquad
%\\
   \displaystyle X_6 = c \sin ( m \s_- )
\end{array}
\end{equation}
where $a=\frac{1}{\sqrt{2}}\cos \gamma,$ $c=\sin \gamma,$
$k =\sqrt{2} m \tan{\gamma}.$
However $k$ is in general arbitrary, so \rf{no} corresponds to the
rigid chiral solutions only when
$$
k =\sqrt{2} m \tan{\gamma} {\rm \ \ is \ integer.}
$$

The expression for the \AdS energy does not changes in $n \to 0$ limit.
The spins transform to (use integral expressions from appendix \ref{appC} and restore $m, n \neq 1$)
\be
\ba{c}
\size
	\J_1=\J_{12} \to 0, \qquad \J_{2}=\J_{34} \to 0, \qquad \J_3 = \J_{56} = mc^2,
\\
\size
   \J_{24}=\left.\frac{\sqrt{2}}{2} m \tan \gamma \cos^2 \gamma \frac{\sin(n \s_+)}{2\pi n}\right|^{2\pi}_0
	\to k a^2
\ea
\ee
in exact agreement with \rf{ESSplane}.

\vspace{20pt}

Being expressed, as harmonic functions with the argument \rf{Fi},
solutions \rf{chan} are not easy to analyze in terms of stability.
Straight forward analysis may be performed only for the rigid chiral solutions
\rf{hhh}, which were proved stable in \ci{art}.
Thus one may also expect \rf{8a} to be stable,
due to their relation to \rf{hhh}.% in the limit $n \to 0.$

\vspace{20pt}

One may hope that generalization of $F_1(\s_+)=\a \cos n \s_+$ to
\be
F_1(\s_+) =	 m_1\s_{+}+ \b_1 \sin (n_1 \s_{+})%\\
\ee
would also give a simple solution. However, in this case, from \rf{sc} we find that
$F_2$ is expressed
via elliptic functions $E, \, F$ and $\Pi$ all together. % (see \rf{f1f2}).
Finding another simple solutions from ansatz \rf{8a} is an open question.

\section*{Summary}

In this paper we have discussed several classical solutions
for a closed bosonic string in the \adss.

First,  we considered small rigid strings with two spins in the \ads \ part of \adss.
Starting from the flat-space solutions \rf{taak} and using
 perturbation theory in the	 curvature of \AdS
space, we constructed  leading terms in the small two-spin solution and
found  corrections to the leading Regge term in
the classical string energy \rf{Ena} and \rf{En2a}.
 We uncovered a discontinuity
in the spectrum of classical strings with equal and unequal winding numbers
in the $Y_1Y_2$ and $Y_3Y_4$ planes ($n_1$ and $n_2$).
In the limit $n_1=n_2$ the expression for 
$E_{n_1 \neq n_2}(S_1,S_2;\l)$ does not coincide with
$E_{n_1 = n_2}(S_1,S_2;\l).$
We then investigated the connection between
small-spin  (flat-space) and large-spin limits of
two-spin string solutions in \ads.
For the $\w_1 = \w_2$ (i.e. $S_1 = S_2$) case we found that, apart from the
trivial cases of folded and circular strings, the general
rigid solution with $S_1=S_2$ in \AdS admitting the large-spin limit does not have
a  small-spin limit.

In the	second part of the paper
 we constructed a new class of
 chiral solutions in \RS \ for which the embedding coordinates of $S^5$ satisfy the
 linear	 Laplace  equations \rf{cheq}. We used the ansatz \rf{chan}
 and obtained the general solution for it in the form \rf{csol2}.
 These solutions
 generalize the previously studied rigid string chiral solutions \rf{hhh} \cite{art}.
 We studied in detail a simple nontrivial	
  example of these solutions \rf{8a}.

There are a number of open questions that we leave
for future investigation. It is of interest to find solutions in
full \ads which correspond to more general flat-space solutions than rigid and
folded ones. The relation between small-spin  and large-spin limits
should be clarified. So far, it  
looks plausible that there is no connection between them 
apart from the trivial limits. 
The origin of the discontinuity in the spectrum
of small-string solutions with $n_1 = n_1$ and $n_1 \neq n_1$
is also not quite clear. 
 Another direction is to study possible 
 applications of chiral solutions \rf{csol2}.

\section*{Acknowledgments}

I am grateful to Arkady Tseytlin for many helpful discussions and
consultations 
and Alin Tirziu for very useful remarks and communications.
This work is supported by a grant of the
Dynasty Foundation and in part by Scientific Schools SS---4142.2010.2.

\appendix

%%%%%%%%%%%%%%%%%%%%%%%%%%%%%%%%%%%%%%%%%%%%
\section{Appendix: Circular and folded strings in \AdS}\label{AppA}
%%%%%%%%%%%%%%%%%%%%%%%%%%%%%%%%%%%%%%%%%%%%
%\refstepcounter{section}
\def\theequation{A.\arabic{equation}}
\setcounter{equation}{0}

%%%%%%%%%%%%%%%%%%%%%%%%%%%%%%%%%%%%%%%%%%%%
\subsection{ The circular string solution}
%%%%%%%%%%%%%%%%%%%%%%%%%%%%%%%%%%%%%%%%%%%%

A particular simple solution of equations \rf{2}, \rf{3}, \rf{6} is
the rigid  circular rotating string \ci{ft2,art,ptt}:
\be
  Y_{05} = \frac{\sqrt{ m^2 +  w^2}}{\sqrt 2 m}\,e^{i\k\t},\;\;\ \ \
  Y_{12} = \frac{\kappa}{2 m} \,e^{i \w\t+im\s},\;\;\ \
  Y_{34} = \frac{\kappa}{2 m} \,e^{i \w\t-im\s},  \label{crot}
\ee
where $w=\sqrt{m^2+\kappa^2}.$
It can also be rewritten in the form
\begin{equation}
\label{circ}
\size \td Y_{05}=\sqrt{1+r^2} e^{i \k \t}, \qquad
\size \td Y_{12}=r \cos(m\s) e^{i \w \t}, \qquad
\size \td Y_{34}=r \sin(m\s) e^{i \w \t},
\end{equation}
where $\w = m \sqrt{1 + 2 r^2}$ and $r = \sinh \r_0 = \frac{\k}{\sqrt 2 m}$
is a radius of the string.
This  is a consistent closed-string solution  periodic in
$O \leq \s < 2 \pi.$

The two spins of the string are equal $\S_1=\S_2=\S$
and are related to the energy by
\be
\mathcal{E}=\kappa+\frac{2 \kappa
\mathcal{S}}{\sqrt{\kappa^2+m^2}}\ ,\ \ \ \ \ \ \
\mathcal{S}=  {\frac{\kappa^2}{ 4 m^2}}	 \sqrt{m^2+\kappa^2} \ .
\ee

In the small-string limit ($\S \to 0$)
the profile of the string reads
\begin{equation}
\size Y_{05} \approx (1+\frac{1}{2} \ \e^2 \ a^2) \ e^{i \sqrt{2} \ \e \ a m \t}, \qquad
\size Y_{12} \approx a \cos(\s) \ e^{i m (1+ \e^2 a^2) \ \t}, \qquad
\size Y_{34} \approx a \sin(\s) \ e^{i m (1+ \e^2 a^2) \ \t}.
\end{equation}
The expression for the classical energy in this limit is
\be\la{A5}
	\E=2\sqrt{m\mathcal{S}}\left( 1 + \frac{\S}{m} + O(\S^2) \right) \qquad {\rm or} \qquad
	E=2\sqrt{ m \sqrt{\l} S } \left( 1+ \frac{S}{m \sqrt{\l}} + O(\l^{-1}) \right).
\ee
Here the classical energy contains nontrivial curvature
corrections which modify the lead\-ing-order flat-space Regge behavior.

%%%%%%%%%%%%%%%%%%%%%%%%%%%%%%%%%%%%%%%%%
\subsection{Folded string solution}
%%%%%%%%%%%%%%%%%%%%%%%%%%%%%%%%%%%%%%%%%

Another simple solution of equations \rf{2}, \rf{3}, \rf{6} is
the classical solution for the
folded	string spinning in the $AdS_3$ part of $AdS_5$
$$
ds^2= -\cosh^2 \rho\ dt^2 + d \rho^2 + \sinh^2 \rho\ d \phi^2
$$
described by \cite{gkp,TirTs2}
\begin{equation}\la{so}
 t= \kappa \tau, \quad \phi= w \tau, \quad \rho=\rho(\sigma), \end{equation}
where
\begin{equation}
\rho'^2 = \kappa^2 \cosh^2 \rho - w^2 \sinh^2 \rho \ .	 \label{snh}
\end{equation}
$\rho$ varies from $0$ to its maximal  value $\rho_*$
\begin{equation}
\coth^2 \rho_* = \frac{w^2}{\kappa^2}\equiv 1+ \frac{1}{l^2} \ .
\end{equation}
Thus $l$ measures the length of the string.
The solution of the	 differential equation (\ref{snh}), i.e.
\begin{equation}
\rho' = \pm \kappa \sqrt{1-l^{-2} {\sinh^2 \rho}}\ , \quad \quad \rho(0)=0
\end{equation}
 can be written in terms of the Jacobi function ${\rm sn}$
\begin{equation}
\sinh \rho=l \ {\rm sn}({\kappa l^{-1}	\sigma},\ -l^2) \ . \label{mlq}
\end{equation}
The periodicity in $\sigma$ implies the following condition on the parameters \ci{gkp}
\begin{equation}
\kappa=l \	_2 F_1(\frac{1}{2},\frac{1}{2};1;-l^2) \la{uu}	\ .
\end{equation}
The classical energy $E=\sqrt{\lambda} \E$ and the spin $S=\sqrt{\lambda}\S$ are found to be
\begin{equation}
\mathcal{E}=l  \ _2 F_1(-\frac{1}{2},\frac{1}{2};1;-l^2),
\ \ \  \qquad \mathcal{S}=
\frac{l^2}{2}\sqrt{1+{l^2}} \ _2F_1(\frac{1}{2},\frac{3}{2};2;-l^2) \ .	 \label{qdr}
\end{equation}
Here we will
be	 interested
 in the short string limit $0 < \epsilon \ll 1,$ $l=a \e$	in which
\begin{equation}
\rho_*=a \e	  - {\frac{1}{6}}\e^3 a^3 +	 O(\e^{5})	\ .
\end{equation}
In the strict limit
$a=0$  or $\k=0$ we get	  $\rho=\rho_*=0$,	so that
the	 string	 shrinks to a point	  with $E=0$.

From  (\ref{qdr}) in the $\epsilon \ll 1$ or
the small  $\mathcal{S}$ limit we obtain
\bea\la{A14}
\mathcal{E}=\sqrt{2 \mathcal{S}}\left(1+\frac{3}{8}\mathcal{S}+O(\S^2)\right),
\label{mfj}
\eea
so the short string limit corresponds to $\S  \ll 1$ and
 the expansion of the  energy looks like
\begin{equation}\la{fj}
E= \sqrt{2 \sqrt{\l} S}\left(1+\frac{3}{8 \sqrt{\l} } S+O(\l^{-1})\right) .
\end{equation}
Expanding the exact solution  (\ref{mlq}) in powers of $\epsilon$ we obtain
\begin{equation}\label{foldR1}
\sinh \r = \e \ a \ \sin \s - \e^3 \ \frac{a^3}{16} (\sin (3\s) +
\sin \s) +O(\e^5)\\
\end{equation}
or equivalently, changing phase $\s \rightarrow \frac{\pi}{2}-\s$
\begin{equation}\label{foldR2}
	\sinh \r = \e \ a \ \cos \s - \e^3 \ \frac{a^3}{16} (-\cos (3\s) + \cos \s) +O(\e^5).
\end{equation}
For the frequencies we have
\be
			\size \w=1+\e^2\ \frac{a^2}{4}+O(\e^4), \qquad	\k=\e \ a - \e^3 \ \frac{1}{4} \ a^3+O(\e^4).
\ee

\section{Appendix: Folded string displaced from the \AdS center $(n_2=0).$}

\def\theequation{B.\arabic{equation}}
\setcounter{equation}{0}

The possibility omitted in section \ref{Sec3} is when one of the
frequencies of the original flat-space solutions ($n_i$) is zero,
while the ``amplitude'' $y_i=\const \neq 0$\foot{One may also consider perturbations under
a flat-space solution with $n_1=n_2=0,$ i.e. a point-like string displaced from the center of \ads.
There are no closed-string solutions in this limit.
}.
We will look for the solutions of (\ref{Car1}), (\ref{Car2}) in the form:
\begin{equation}\label{n20}
	\ba{ll}
	\size
	y_1(\s) = \e \ a \sin(\s n)+ \e^3 z_1(\s)\\
	\size
	y_2(\s)= \e \ b + \e^3 z_2(\s),
	\ea
\end{equation}
where $n \in Z$ and
\be\label{wwk4}
\ba{ll}
	\size
	\w_1= n (1 + \e^2 \ \tilde{\w}_1), \qquad \
	\size
	\w_2= \e \ \tilde{\w}_2	 \\
	\size
	\k=\e \ \k_0 + \e^3 \ \k_1, \quad \qquad \k_0^2 = a^2 n^2.
\ea
\ee
It follows from \rf{dsexpC}, that expansion of $\w_i^2$ must consist of the
even powers of $\e.$ So if $n_2 = 0$ the leading order of $\w_2$ is $\e.$

From (\ref{Car1}), (\ref{Car2}) one obtains the set of equations:
\be
&&	\size - b \ (n^2 z_1 + z_1'') + a \sin(n \s) \ z_2'' = 2 a b \sin(n \s) (  \tilde{\w}_1 \ n^2
	- \tilde{\w}_2^2) \label{sys4a}
\\[6pt]
&& \ba{ll}	\size 2 a n \ [ \sin(n \s ) n z_1 + \cos(n \s) z_1'] = 2 \chi - b^2 \tilde{\w}_2^2 + a^2 b^2 n^2 \\[6pt]
	\qquad \qquad \qquad- 2 a^2 \ \tilde{\w}_1 \ n^2 \sin^2(n \s)  +2 a^4 n^2 \sin^2(n\s)  - a^4 n^2 \sin^4(n\s).
	\ea \label{sys4b}
\ee
Here $\chi^2=\k_1^2 \k_0^2.$
These system can be readily solved. The solution of \rf{sys4b} is straight forward:
\begin{equation}
\ba{ll}
	\size z_1= C_1 \cos(n \s) +a n \ \s \ \cos(n \s) \left(\tilde{\w}_1-\frac{1}{4} a^2\right)
	+ \frac{\sin(n \s)}{2 a n^2} (2 \chi - b^2 \tilde{\w}_2^2) \\
\qquad \qquad \qquad \qquad \size - a \sin(n \s) \left( \tilde{\w}_1 - \frac{1}{2}(a^2+ b^2) \right)
	 -\frac{1}{4} a^3 \sin^2(n \s) \cos(n \s).
\ea
\end{equation}
Employing the closed-string periodicity condition \rf{6}, one finds
\begin{equation}
	\tilde{\w}_1  = \frac{a^2}{4}.
\end{equation}
Then the solution for $z_2$ is
\begin{equation}
	z_2 = C_2+\s C_3-\frac{1}{4} a^2 b \cos(2 n \s)+\s^2 \frac{b}{2} \left(a^2 n^2-
	\tilde{\w}_2^2\right).
\end{equation}
Making use of the closed-string periodicity conditions, one finds
\begin{equation}
	\tilde{\w}_2= \pm a n, \qquad C_3=0 \ .
\end{equation}
There is no additional constraints on the parameters $C_1, C_2, \chi,$ so
the solution of \rf{sys4a}, \rf{sys4b} is
\be\label{zz4}
\ba{c}
\size z_1= C_1\cos(n \s) +\frac{1}{16} a^3 (3 \sin(n \s)-\sin(3 n \s))+\frac{\k_1 \sin(n \s)}{n}
\\
\size z_2=C_2-\frac{1}{4} a^2 b \cos(2 n \s)
\\
	\size
	\w_1= n (1 + \e^2 \ \frac{a^2}{4}), \qquad \
	\size
	\w_2= \pm \e \ a n, \qquad
	\size
	\k=\e \ a n + \e^3 \ \k_1.
\ea
\ee
It is not hard to see, that due to $ \k \approx \w_2,$ non-Cartan components of the spin
$\S_{0i}$ do not vanish. This solution can be rotated by boost to
a folded string one.

\section{Appendix: Spins for the Sine-cosine solutions}\label{appC}

\def\theequation{C.\arabic{equation}}
\setcounter{equation}{0}

In this section we will calculate the components of spin
$\J_{ij}=\int \limits^{2\pi}_0 \frac{d \sigma}{2\pi}[X_i
\dot{X}_j-X_j \dot{X}_i]$ for the Sine-cosine solutions \rf{8a}.
Set for the simplicity $n=m_3=1.$

\noindent Cartan components of the spin are
\begin{equation}\label{S12}
\begin{array}{ll}
   \displaystyle
   \J_1=\J_{12}= \int \limits^{2\pi}_0 \frac{d \sigma}{2\pi}[X_1 \dot{X}_2-X_2 \dot{X}_1] \\
   \qquad \size =\frac{1}{4}\sin(2\gamma)\int \limits^{2\pi}_0 \frac{d \sigma}{2\pi} \left[
   \sin(\tau+\sigma+\pi/4)\sin(2 \tan \gamma \sin(\tau+\sigma+\pi/4))\right.\\
	\qquad \qquad \qquad \qquad \qquad \qquad \size	 +
   \left.\sin(\tau+\sigma-\pi/4)\sin(2 \tan \gamma \sin(\tau+\sigma-\pi/4))
	\right]\\
   \displaystyle
   \qquad =\frac{1}{2}\sin(2\gamma)\int \limits^{2\pi}_0 \frac{d \zeta}{2\pi}
   \sin(\zeta)\sin(2 \tan \gamma \sin(\zeta))
   =\frac{1}{2}\sin(2\gamma) { \rm BesselJ}_1(2 \tan \gamma)
\end{array}
\end{equation}
(see Appendix \ref{AppBessel} for a proof of the equality on the last line);
\begin{equation}\label{S34}
\begin{array}{ll}
   \displaystyle
   \J_2=\J_{34}= \int \limits^{2\pi}_0 \frac{d \sigma}{2\pi}[X_3 \dot{X}_4-X_4 \dot{X}_3] \\
   \displaystyle
   \qquad \size =-\frac{1}{4}\sin(2\gamma)\int \limits^{2\pi}_0 \frac{d \sigma}{2\pi} \left[
   \sin(\tau+\sigma+\pi/4)\sin(2 \tan \gamma \sin(\tau+\sigma+\pi/4))\right.\\
   \qquad \qquad \qquad \qquad \qquad \qquad \size	  -
   \left.\sin(\tau+\sigma-\pi/4)\sin(2 \tan \gamma \sin(\tau+\sigma-\pi/4))
	\right]\\
	\displaystyle
	\qquad =\frac{1}{4}\sin(2\gamma)\int \limits^{2\pi}_0 \frac{d \zeta}{2\pi}
   \left[\sin(\zeta)\sin(2 \tan \gamma \sin(\zeta))-\sin(\zeta)\sin(2 \tan \gamma \sin(\zeta))\right]
	=0 \ ;
\end{array}
\end{equation}
\begin{equation}\label{S56}
   \J_3=\J_{56}= \int \limits^{2\pi}_0 \frac{d \sigma}{2\pi}[X_5 \dot{X}_6-X_6 \dot{X}_5]=\sin^2 \gamma \ .
\end{equation}
Non-Cartan components of the spin are
\begin{equation}\label{S13}
   \J_{13}= \int \limits^{2\pi}_0 \frac{d \sigma}{2\pi}[X_1 \dot{X}_3-X_3 \dot{X}_1]=
   \frac{1}{2\sqrt{2}}\sin(2\gamma) \int \limits^{2\pi}_0 \frac{d \sigma}{2\pi}\sin(\tau+\sigma)=0 \ ;
\end{equation}
\begin{equation}\label{S24}
   \J_{24}= \int \limits^{2\pi}_0 \frac{d \sigma}{2\pi}[X_2 \dot{X}_4-X_4 \dot{X}_2]=
   -\frac{1}{2\sqrt{2}}\sin(2\gamma) \int \limits^{2\pi}_0 \frac{d \sigma}{2\pi}\cos(\tau+\sigma)=0 \ ;
\end{equation}
\begin{equation}\label{S14}
\begin{array}{ll}
   \displaystyle \J_{14}= \int \limits^{2\pi}_0 \frac{d \sigma}{2\pi}[X_1 \dot{X}_4-X_4 \dot{X}_1]\\
   \displaystyle
   \qquad = \frac{1}{2\sqrt{2}}\sin(2\gamma) \int \limits^{2\pi}_0 \frac{d \sigma}{2\pi}
   \left[
   \sin(\tau+\sigma-\pi/4)\cos(2 \tan \gamma \sin(\tau+\sigma-\pi/4))\right.\\
   \qquad \qquad \qquad \qquad \qquad \qquad \size
   +\left.
   \sin(\tau+\sigma+\pi/4)\cos(2 \tan \gamma \sin(\tau+\sigma+\pi/4))
   \right]\\
   \qquad \size
   =\frac{1}{\sqrt{2}}\sin(2\gamma) \int \limits^{2\pi}_0 \frac{d \zeta}{2\pi}
   \sin(\zeta)\cos(2 \tan \gamma \sin(\zeta))\\
   \qquad \displaystyle
   =\frac{1}{\sqrt{2}}\sin(2\gamma) \int \limits^{2\pi}_0 \frac{d \zeta}{2\pi}
	\sum\limits^{\infty}_{l=0} \frac{(-1)^l}{(2l)!} (2 \tan \gamma)^{2l} \sin^{2l+1}(\zeta)=0 \ ;
\end{array}
\end{equation}
\begin{equation}\label{S23}
\begin{array}{ll}
   \displaystyle
   \J_{23}= \int \limits^{2\pi}_0 \frac{d \sigma}{2\pi}[X_2 \dot{X}_3-X_3 \dot{X}_2]=\\
   \displaystyle
   \qquad = \frac{1}{2\sqrt{2}}\sin(2\gamma) \int \limits^{2\pi}_0 \frac{d \sigma}{2\pi}
   \left[
   \sin(\tau+\sigma-\pi/4)\cos(2 \tan \gamma \sin(\tau+\sigma-\pi/4))\right.\\
   \qquad \qquad \qquad \qquad \qquad \qquad \size
   - \left.\sin(\tau+\sigma+\pi/4)\cos(2 \tan \gamma \sin(\tau+\sigma+\pi/4))
   \right]\\
   \displaystyle
   \qquad =\frac{1}{2\sqrt{2}}\sin(2\gamma) \int \limits^{2\pi}_0 \frac{d \zeta}{2\pi}
   \left[\sin(\zeta)\cos(2 \tan \gamma \sin(\zeta))-
   \sin(\zeta)\cos(2 \tan \gamma \sin(\zeta))\right]=0 \ .
\end{array}
\end{equation}
Here we used that the integral over the period from odd powers of 
sine or cosine is zero \cite{grr}:
\begin{equation}\la{sck}
   \int \limits^{2\pi}_0 d\zeta \sin^{2l+1} \zeta=0, \qquad \int \limits^{2\pi}_0 d\zeta \cos^{2l+1} \zeta=0.
\end{equation}

To prove that $\J_{5j}=\J_{6j}=0,$ $j=1,2,3,4,$
consider the following expansion of $X_{ij}:$
\begin{equation}\label{xijexp}
   X_{12}=\sum \limits^\infty_{l=0}g^{(1)}_l e^{il(\sigma+\tau)}, \qquad
   X_{34}=\sum \limits^\infty_{l=0}g^{(2)}_l e^{il(\sigma+\tau)}, \qquad
   X_{56}=\sum \limits^\infty_{l=0}h_l e^{il(\sigma-\tau)}.
\end{equation}
One can show that ``cross-spins'' (non-Cartan components of spins)
between right- and left-chiral waves always vanish,
i.e. for each pair of right- and left-chiral summands in \rf{xijexp}:
\begin{equation}
   Z_1+i Z_2=G e^{in(\sigma+\tau)}, \qquad Z_3+i Z_4=H
   e^{im(\sigma-\tau)}, \qquad n,m ={\rm integer}
\end{equation}
the correspondent contribution ($\J^Z_{ij}$) into $\J_{5j}, \ \J_{6j}, \ j =1,2,3,4$
is zero.

Let us calculate the following values
\begin{equation}
\begin{array}{ll}
   \size {\J}^Z_{+}=\int \limits^{2\pi}_0 \frac{d \sigma}{2\pi}[z_1 \dot{z}_2 - z_2 \dot{z}_1 ]=
   [\J^Z_{13}-\J^Z_{24}]+i[\J^Z_{23}+\J^Z_{14}]\\
\qquad \qquad \size =i\int \limits^{2\pi}_0 \frac{d \sigma}{2\pi}
   GH \ (m-n)e^{i\sigma (n-m)+i\tau (n+m)}=\left\{
   \begin{array}{ll}
	   0, \quad \rm{for} \quad m=n\\
	   0, \quad \rm{for} \quad m \neq n
   \end{array}
	 \right.\\
   \size \J^Z_{-}=\int \limits^{2\pi}_0 \frac{d \sigma}{2\pi}[z_1 \dot{z}^{+}_2 - z^{+}_2 \dot{z}_1 ]=
   [\J^Z_{13}+\J^Z_{24}]+i[\J^Z_{23}-\J^Z_{14}]\\
\qquad \qquad \size =i\int \limits^{2\pi}_0 \frac{d \sigma}{2\pi}
   G H \ (m-n)e^{i\sigma (n-m)+i\tau (n+m)}=\left\{
   \begin{array}{ll}
	   0, \quad \rm{for} \quad m=n\\
	   0, \quad \rm{for} \quad m \neq n.
   \end{array}
	 \right.
\end{array}
\end{equation}
Cross-spins for each left-right chiral pair in the
expansion \rf{xijexp} vanish. We have
\begin{equation}
   \J_{5j}=\J_{6j}=0, \qquad j=1,2,3,4.
\end{equation}

\section{Appendix: Bessel functions}\label{AppBessel}

\def\theequation{D.\arabic{equation}}
\setcounter{equation}{0}

In this section we will prove the relation
\begin{equation}\la{D1}
   {\rm BesselJ}_1 (x)=\int \limits^\pi_{-\pi} \frac{d\alpha}{2\pi} \sin{\alpha} \sin{(x\sin{\alpha})}.
\end{equation}

Two formulas from the theory of the Bessel functions are of use \cite{bw}:
\begin{itemize}
\item
Integral representation of the Bessel functions
\begin{equation}\label{Jn}
   {\rm BesselJ}_n (x)=\int \limits^\pi_{-\pi} \frac{d\alpha}{2\pi}e^{-ix\, \sin\alpha+i n \alpha}.
\end{equation}
\item
Recurrent formula
\begin{equation}\label{JnRec}
   \frac{d}{dx}\left( \frac{{\rm BesselJ}_\nu (x)}{x^\nu} \right)=\frac{{\rm BesselJ}_{\nu+1} (x)}{x^\nu}.
\end{equation}

\end{itemize}

Let us take a derivative from ${\rm BesselJ}_0(x)$ in the integral
representation:
\begin{equation}
\begin{array}{c}
   \size \frac{d}{dx}  {\rm BesselJ}_0 (x)=
   -i\int \limits^\pi_{-\pi} \frac{d\alpha}{2\pi}\sin{\alpha} \ e^{-ix\, \sin\alpha}
   =\int \limits^\pi_{-\pi} \frac{d\alpha}{2\pi}
   \left[ -i\sin{\alpha} \cos{(x\sin{\alpha})}-\sin{\alpha} \sin{(x\sin{\alpha})} \right] .
\end{array}
\end{equation}
The Taylor expansion of $\sin{(x\sin{\alpha})}$ and
$\cos{(x\sin{\alpha})}$ consist of odd and even powers of $\sin{\alpha},$ respectively.
Making use of \rf{sck}, one finds
\begin{equation}
   \frac{d}{dx} {\rm BesselJ}_0 (x)=-\int \limits^\pi_{-\pi} \frac{d\alpha}{2\pi} \sin{\alpha} \sin{(x\sin{\alpha})}.
\end{equation}
Then by employing (\ref{JnRec})
\begin{equation}
   \frac{d}{dx} {\rm BesselJ}_0 (x)=-{\rm BesselJ}_{1} (x)
\end{equation}
and we end up with \rf{D1}.

\end{document}